\documentclass[fleqn,usenatbib]{mnras}
\usepackage{amssymb}
\usepackage{newtxtext,newtxmath}

\usepackage[T1]{fontenc}
\usepackage{rotating}
\makeatletter
\let\@makecaption=\SFB@makefigurecaption
\makeatother
\usepackage{tabularx}

\newcommand{\kms}{km\,s$^{-1}$}
\newcommand{\oii}{O\,{\scriptsize II}}
\newcommand{\oiii}{O\,{\scriptsize III}}

\newcommand{\cii}{C\,{\scriptsize II}}
\newcommand{\ciii}{C\,{\scriptsize III}}
\newcommand{\sii}{S\,{\scriptsize II}}
\newcommand{\nii}{N\,{\scriptsize II}}
\newcommand{\niii}{N\,{\scriptsize III}}
\newcommand{\niv}{N\,{\scriptsize IV}}
\newcommand{\neiii}{Ne\,{\scriptsize III}}
\newcommand{\nv}{N\,{\scriptsize V}}
\newcommand{\civ}{C\,{\scriptsize IV}}

\newcommand{\hii}{H\,{\scriptsize II}}
\newcommand{\hei}{He\,{\scriptsize I}}
\newcommand{\heii}{He\,{\scriptsize II}}

\newcommand{\hbeta}{H$\beta$}

\newcommand{\lya}{Ly$\alpha$}

\DeclareRobustCommand{\VAN}[3]{#2}
\let\VANthebibliography\thebibliography
\def\thebibliography{\DeclareRobustCommand{\VAN}[3]{##3}\VANthebibliography}

\usepackage{graphicx}	
\usepackage{amsmath}	
\usepackage{float}
\usepackage{bm}
\usepackage{booktabs}
\usepackage{orcidlink}
\hypersetup{colorlinks=true, citecolor=blue, linkcolor=blue, urlcolor=blue}

\title[OMEGA: Precise C-N-O abundances at $z\geqslant6$]{Precise C-N-O Abundances of Individual Galaxies in the First Billion Years from Ultra-Deep, Rest-UV-to-Optical JWST Spectroscopy}

\author[Rai et al.]{
Raunaq Singh Rai\,\orcidlink{0009-0000-8489-7211}$^{1}$\thanks{E-mail: raunaq.rai.25@ucl.ac.uk}, 
Guido Roberts-Borsani\,\orcidlink{0000-0002-4140-1367}$^{1}$, 
Ryan Sanders\,\orcidlink{0000-0003-4792-9119}$^{2}$, 
Alice E. Shapley\,\orcidlink{0000-0003-3509-4855}$^{3}$,
\newauthor
Leonardo Clarke\,\orcidlink{0000-0003-1249-6392}$^{3}$, 
Pascal A. Oesch\,\orcidlink{0000-0001-5851-6649}$^{4,5}$, 
Daniel Schaerer\,\orcidlink{0000-0001-7144-7182}$^{4,6}$, 
Rui Marques-Chaves\,\orcidlink{0000-0001-8442-1846}$^{4}$,
\newauthor
Romain A. Meyer\,\orcidlink{0000-0001-5492-4522}$^{4}$,
Emma Giovinazzo\,\orcidlink{0009-0004-3835-0089}$^{4}$,
and Shreya Karthikeyan\,\orcidlink{0009-0002-6186-0293}$^{3}$, 
\\
$^{1}$Department of Physics \& Astronomy, University College London, London, WC1E 6BT, UK\\
$^{2}$Department of Physics \& Astronomy, University of Kentucky, 505 Rose Street, Lexington, KY 40506, USA \\
$^{3}$University of California, Los Angeles, 475 Portola Plaza, Los Angeles, CA, 90095, USA \\
$^{4}$Department of Astronomy, University of Geneva, Chemin Pegasi 51, 1290 Versoix, Switzerland \\
$^{5}$Cosmic DAWN Center, Niels Bohr Institute, University of Copenhagen, Jagtvej 128, K\o benhavn N, DK-2200, Denmark \\
$^{6}$CNRS, IRAP, 14 Avenue E. Belin, 31400 Toulouse, France
}

\date{Accepted XXX. Received YYY; in original form ZZZ}

\pubyear{\the\year{}}

\begin{document}
\label{firstpage}
\pagerange{\pageref{firstpage}--\pageref{lastpage}}
\maketitle

\begin{abstract}

We present ultra-deep JWST/NIRSpec $R\sim1000$ spectroscopy of nine star-forming galaxies at $z=6.0-8.3$ in the GOODS-N field, drawn from the Origins of Metal Enrichment in Galaxies at cosmic dAwn (OMEGA) survey. Paired with ancillary ultra-deep G140M observations, the G395M integrations from OMEGA yield continuous rest-frame UV-to-optical coverage and the full suite of diagnostics needed for precise characterisations and measures of the chemical abundances of these systems. Crucially, the auroral [\oiii]$\lambda$4363 line is detected in every galaxy, enabling the determination of electron temperatures and direct oxygen abundances of $12+\log(\mathrm{O/H})=7.14-8.18$, independent of strong-line calibrations that dominate systematics at these redshifts. We couple direct oxygen abundances with measurements of rest-frame UV emission-line strengths and ratios to infer the physical properties of our sample. Most importantly, we trace the relationship between oxygen abundance and both C/O and N/O abundance ratios with high fidelity,  demonstrating robustly that $z=6.0-8.3$ galaxies follow the local trend in C/O vs. O/H, and that inferences regarding N/O are sensitive to the choice of rest-frame UV (\niv] or \niii]) or optical ([\nii]) tracers. This pilot study demonstrates the power of combining ultra-deep JWST/NIRSpec $R\sim1000$ rest-frame UV and optical spectroscopy, which now requires statistical samples to uncover the nebular properties of the full population of $z\geq 6$ star-forming galaxies. 

\end{abstract}




\section{Introduction}
The chemical enrichment histories of galaxies in the first billion years represent fundamental tracers of early galaxy formation and evolution, as well as probes of the processes that synthesised the first heavy elements.
In particular, carbon (C), nitrogen (N) and oxygen (O) are synthesised through different mechanisms and dispersed into the interstellar medium (ISM) on different timescales. Their relative abundances therefore serve as key observational probes of the enrichment processes regulating the growth of early star-forming systems. Carbon is produced via triple-$\alpha$ reactions in both massive stars ($>10\,M_\odot$) and low-mass AGB stars ($\simeq1-4\,M_\odot$), and oxygen almost entirely from the core-collapse supernovae of massive stars. Nitrogen is made in both massive and intermediate-mass stars ($\simeq4-7\,M_\odot$) depending overall metallicity. Carbon and oxygen are typically released rapidly on timescales of a few tens of Myr, while nitrogen enrichment from AGB stars generally occurs on longer timescales of $\sim100-300$ Myr. Considering the abundances of C, N, and O, in concert sheds light on the relative importance of different stellar evolution processes operating in star-forming galaxies \citep[e.g.,][]{berg2019,kobayashi2020}, particularly at the earliest cosmic times.

The James Webb Space Telescope (JWST) has pushed C-N-O and chemical abundance measurements out to unprecedented redshifts, enabling characterisation of galaxy metallicities and early enrichment histories  within only a few hundred Myr of the Big Bang (e.g., \citealt{naidu2026,castellano2024,deugenio24,robertsborsani2024,robertsborsani2026}). To this end, the near omnipresence of oxygen and Balmer emission lines from large samples of NIRSpec prism spectra has demonstrated rapid and significant oxygen enrichment in large samples of $6\leqslant z \leqslant10$ galaxies, with clear evolution seen on average at $M_*\sim 10^8 M_{\odot}$ from more metal-poor ($\sim$0.05\,$Z_{\odot}$) sources at $z\sim10$ to more enriched ($\sim0.15$\,$Z_{\odot}$) systems by $z\sim6$ \citep{robertsborsani2024,heintz2025}. The scatter is large, however, with a significant number of sources at any given epoch revealing significantly more pristine \citep[$<0.05$\,$Z_{\odot}$;][]{cullen2025} or enriched \citep[$\sim0.50$\,$Z_{\odot}$;][]{rowland2026} gas reservoirs relative to the mean, and a number of remarkable individual systems with metallicities as high as solar by $z\sim7$ \citep{shapley25}. MIRI spectroscopy has extended such measurements further still, beyond the $z\sim10$ frontier, albeit it over much smaller samples. Those observations suggest a shift towards even lower 
metallicities and a gradual approach towards chemically pristine systems \citep{hsiao24,zavala25,marqueschaves26_metal,alvarezmarquez25,alvarezmarquez26}.

Arguably the greatest surprise, however, has come from the combination of rest-frame UV and optical spectroscopy, and pairing oxygen abundances to those of nitrogen and carbon \citep[e.g.,][]{topping2024,topping2025,senchyna2024,morel2026}. Measurements of supersolar N/O ratios in some particularly luminous $z>9$ sources have indicated a nitrogen enhancement relative to $z\sim0$ observations that is challenging to explain based on AGB N-enrichment alone, due to the timescales involved and short lifetimes of their hosts \citep[e.g.,][]{cameron2023,castellano2024}. Together with measurements of subsolar C/O ratios \citep{jones2023,Ji2025}, the implications of such observations are profound given the dramatic nature of some solutions invoked to generate greater N yields, including the presence of short-lived Wolf-Rayet stars \citep[e.g.][]{berg2026,kobayashi2024}, rare very-to-super massive stars (VMS and SMS, respectively; e.g., \citealt{marqueschaves2026,charbonnel2023}), and even a variable (top-heavy) initial mass function (IMF) \citep[e.g][]{nagele2023, cameron2024}. Both individually and collectively, these measurements challenge our understanding of the origins and timescales of chemical evolution and star formation at high redshift.

Despite these remarkable observations and discoveries, however, statistics remain extremely poor and the majority of existing measurements are derived from emission line signatures in low-resolution prism spectroscopy that are susceptible to both sample and instrumental biases. Specifically, oxygen abundances based on the ratios of strong rest-optical emission lines can suffer up to $\sim1$ dex scatter and are calibrated primarily to $z\sim2-3$ samples \citep{sanders2024,sanders26}. At the same time, N and C abundances come primarily from high-ionisation ($\sim24-48$ eV) \niv], \niii], \civ, and \ciii] lines in the rest-frame UV that are particularly affected by poor spectral resolution and limited depth (\citealt{robertsborsani2024,robertsborsani2026}, but see \citealt{topping2024}), while neglecting important contributions from low-ionisation gas (e.g., [\nii]; \citealt{schaerer2026}). Compounding issues further, interpretation of these measurements rely on crude and untested assumptions (rather than direct measurements) of the state of the ISM (e.g., gas density and temperature, ionisation conditions) and underlying stellar populations, predominantly due to the lack of suitable tracers. As such, the accuracy of the vast majority of current abundance measurements remains questionable \citep{martinez2025,arellanocordova2026}, while the prevalence of high-ionisation lines, their resulting nitrogen and carbon abundances, and the origins of anomalous ratios has yet to be determined.

NIRSpec $R\sim1000$ spectroscopy offers a compelling alternative to prism spectroscopy,  providing the ideal trade-off between spectral resolution and sensitivity and allowing for robust measurements of faint and closely spaced emission lines (or doublets) that yield the electron density and temperature measurements required for gold-standard chemical abundances and ionisation zone corrections. Spectral resolution and sensitivity are especially essential for measurements of the auroral [\oiii]$\lambda$4363 \AA\ line. This feature, together with strong [\oiii]$\lambda\lambda$4960,5008 \AA, serves as the standout diagnostic with which to deliver robust electron temperatures \citep[e.g.,][]{sanders2024,sanders2016te,curti2023,morishita24,laseter2024}. These temperatures in turn anchor accurate metallicity, N/O and C/O measurements, and the interpretation of the rich suite of interstellar and stellar emission and absorption lines in the rest-UV. Despite these considerations, however, observations of all but the brightest individual $z>6$ sources \citep[e.g., ][]{topping2024,topping2025,chen2026} typically lack the required depth for robust constraints. To circumvent these issues, a number of recent studies have leveraged the constraining power of existing samples together with stacking techniques, to probe the chemical abundances of $z\geqslant6$ galaxies through high signal-to-noise (S/N) $R\sim1000$ composite spectra \citep{rai2026,umeda2026,rusakov2026}. For instance, in a recent example, \citet{rai2026} investigated the prevalence and degree of nitrogen enhancement in stacks of 135 star-forming sources at $z>6$, finding weak but ubiquitous \niv]\ emission and supersolar N/O ratios with links to recent star formation histories that point to significant AGB enrichment. While such stacking efforts go a long way towards pinpointing the prevalence and origins of C-N-O enrichment, recent studies using ultra-deep $R\sim1000$ spectroscopy have shown the importance of abundance constraints in individual objects, many of which may be short-lived and diluted in composite spectra (e.g., \citealt{berg2026,arellanocordova2025}).

Here we present new, ultra-deep $R\sim1000$ NIRSpec observations and chemical characterisations of a sample of individual $z\simeq6.0-8.3$ star-forming galaxies. Our analysis showcases the constraining power and potential of combining NIRSpec rest-frame UV and optical spectroscopy (i.e., G140M+G395M)  over larger samples to accurately constrain the chemical evolution of galaxies in the first billion years. The paper is organised as follows. The observations, reduction procedure, and measurements are described in Section~\ref{sec:sample} and Section~\ref{sec:methods}. The sample demographics are described in Section~\ref{sec:demographics}, and metallicities and abundance patterns are presented in Section~\ref{sec:metals}. We discuss the implications of these results in Section~\ref{sec:discussion}, and conclude in Section~\ref{sec:summary}.

Throughout this paper we adopt the flat $\Lambda$CDM cosmology of \citet{planck2018}, with $H_{0} = 67.7$~km~s$^{-1}$~Mpc$^{-1}$, $\Omega_{\rm m} = 0.31$ and $\Omega_{\Lambda} = 0.69$. All magnitudes are quoted in the AB system \citep{oke1983}. Elemental abundances are reported on the standard logarithmic scale, with the solar reference values of \citet{asplund2009} ($12 +\log(\mathrm{O}/\mathrm{H})_{\odot} = 8.69$, $\log(\mathrm{N}/\mathrm{O})_{\odot} = -0.86$, and $\log(\mathrm{C}/\mathrm{O})_{\odot} = -0.26$). Unless stated otherwise, we use metallicity to refer to the gas-phase oxygen abundance, $12 + \log(\mathrm{O}/\mathrm{H})$.

\section{Observations and data reduction}
\label{sec:sample}

\subsection{The OMEGA survey}
\label{sec:programme}
The Origins of Metal Enrichment in Galaxies at cosmic dAwn (OMEGA) program is a JWST Cycle 4 extragalactic survey (GO 7729, PI Roberts-Borsani) designed to obtain precise abundance measurements in high-redshift ($z\gtrsim6$) galaxies through especially deep NIRSpec spectroscopy. With a total of 67.2 hours observing time in G395M (affording $\sim2.75-5.40 \mu$m wavelength coverage), distributed over 4 pointings, the primary goal of the $R\sim1000$ observations is to detect (or place stringent upper limits on) the electron temperature-sensitive [\oiii]$\lambda$4363 \AA\ emission line. This feature typically remains undetected in individual $R\sim1000$ and prism observations due, respectively, to shallow depth and poor spectral resolution (causing blending with nearby H$\gamma$ emission). Combined with strong [\oiii]$\lambda\lambda$4960,5008 \AA, such measurements provide the gold-standard oxygen abundances that anchor metallicity measurements and are currently lacking in the vast majority of confirmed high-redshift galaxies.

A detailed description of the survey design and observations will be provided in a forthcoming survey paper (Roberts-Borsani et al., in prep). However, as a brief overview, the OMEGA observations were split into four non-overlapping multi-shutter array (MSA) masks across the two GOODS fields (two in GOODS-N and two in GOODS-S) each observed for $47{,}268$\,s ($13.1$\,hours) of on-source integration with the G395M/F290LP grating, corresponding to $16.8$\,hours of charged time per mask. The primary $z\geqslant6$ targets were selected based on existing strong-line spectroscopic measurements (e.g., \citealt{robertsborsani2024,scholtz26,meyer24}), allowing for electron temperature predictions for [\oiii]$\lambda$4363 \AA, while affording the full suite of required diagnostics of accurate O/H abundance measurements. Within those masks, a number of primary targets already benefitting from deep rest-frame UV ancillary spectroscopy were given higher priority, so as to extend the effective wavelength coverage down to $\sim1 \mu$m and provide the full suite of tracers required to accurately map out the ISM, stellar, and C-N-O abundances of $z>6$ galaxies. Here we present a subset of the full survey, taken over the two GOODS-N pointings.

\subsection{Ancillary spectroscopy}
\label{sec:ancillary}

For sources with clearly detected ($>3\sigma$) [\oiii]$\lambda$4363 \AA\ and without obvious AGN signatures in the OMEGA spectra, we verify the availability of complementary rest-UV spectroscopy from deep G140M (F070LP or F100LP) observations, in order to place strong constraints on C and N abundances and their ratios with OMEGA-derived O/H metallicities. We draw the rest-UV observations from the SPURS (GO 9214, PIs Mason \& Stark; \citealt{spurs}), DIVER (GO 8018, PI Lin; \citealt{lin2025}), AURORA (GO 1914, PIs Shapley \& Sanders; \citealt{aurora}), JADES (GTO 1181, PI Eisenstein; \citealt{jades1}), and GO 4762 (PI Fujimoto) surveys. SPURS features 29.2, 7.9, and 2.9 hours in G140M/F100LP, G235M/F170LP, and G395M/F290LP, respectively; DIVER is designed with 20.4 hours in G140M/F070LP; AURORA observed for 12.2, 8.0, and 4.2 hours in G140M/F100LP, G235M/F170LP, and G395M/F290LP, respectively; JADES medium-resolution spectroscopy features a range of exposure times in G140M/F070LP, G235M/F170LP, and G395M/F290LP, and here we fold in 0.85-hour integrations in G140M/F070LP and G395M/F290LP; finally GO-4762 has 2.1 hours in both G140M/F100LP and G395M/F290LP. As a minimum, we further require clear detections of the \ciii]$\lambda\lambda$1907,1909 \AA\ emission line, for reliable C/O abundances. Nine sources satisfy our selection criteria, which we list in Table~\ref{tab:sample} along with their spectroscopic redshifts and a number of global and ISM properties (see ensuing sections), and whose NIRCam morphologies and micro-shutter coverage are shown in Figure~\ref{fig:cutouts}.

\begin{table*}
\centering
\renewcommand{\arraystretch}{1.3}
\begin{tabular}{lcccccccc}
\toprule
OMEGA & $z$ & $M_{\rm UV}$ & $\log(M_\star/\mathrm{M}_\odot)$ & $12+\log(\mathrm{O/H})$ & $T_{\rm e}(\mathrm{O\,\textsc{iii}}\,\lambda4363)$ & $O32$ & $R3$ & $\log\mathrm{SFR}(\mathrm{H}\alpha)$ \\
MSAID & & (mag) & & (direct, $\lambda4363$) & ($10^4$\,K) & & & ($\mathrm{M}_\odot\,\mathrm{yr}^{-1}$) \\
\midrule
1   & 8.27889 & $-19.56^{+0.15}_{-0.14}$ & $8.19^{+0.53}_{-0.14}$ & $7.54^{+0.06}_{-0.05}$ & $2.30 \pm 0.15$ & $>77$ & $8.20^{+0.41}_{-0.38}$ & $0.91^{+0.17}_{-0.16}$\,$^{\star}$ \\
6   & 6.80833 & $-20.22^{+0.04}_{-0.04}$ & $8.75^{+0.42}_{-0.14}$ & $7.98^{+0.06}_{-0.07}$ & $1.46 \pm 0.09$ & $8.5^{+0.8}_{-1.2}$ & $7.12^{+0.17}_{-0.18}$ & $0.82^{+0.02}_{-0.02}$ \\
22  & 6.31276 & $-20.17^{+0.04}_{-0.04}$ & $8.24^{+0.60}_{-0.19}$ & $7.46^{+0.08}_{-0.08}$ & $1.97 \pm 0.17$ & $>2.5$ & $5.13^{+0.10}_{-0.10}$ & $0.94^{+0.01}_{-0.01}$ \\
24  & 6.04756 & $-18.69^{+0.14}_{-0.12}$ & $7.92^{+0.78}_{-0.23}$ & $7.68^{+0.13}_{-0.12}$ & $1.97 \pm 0.29$ & -- & $8.15^{+0.39}_{-0.53}$ & $0.74^{+0.29}_{-0.02}$ \\
75  & 7.20485 & $-19.88^{+0.11}_{-0.10}$ & $7.89^{+0.92}_{-0.04}$ & $7.85^{+0.12}_{-0.10}$ & $1.56 \pm 0.15$ & $14.2^{+1.8}_{-1.9}$ & $6.76^{+0.31}_{-0.29}$ & $0.68^{+0.03}_{-0.02}$ \\
76  & 7.20238 & $-21.07^{+0.03}_{-0.03}$ & $8.44^{+0.31}_{-0.51}$ & $7.93^{+0.18}_{-0.15}$ & $1.52 \pm 0.23$ & $14.0^{+2.8}_{-2.0}$ & $7.51^{+0.57}_{-0.56}$ & $0.59^{+0.25}_{-0.05}$ \\
77  & 7.18922 & $-20.97^{+0.10}_{-0.09}$ & $8.65^{+0.46}_{-0.18}$ & $7.91^{+0.10}_{-0.09}$ & $1.57 \pm 0.14$ & $12.9^{+1.3}_{-0.9}$ & $7.83^{+0.30}_{-0.27}$ & $0.70^{+0.03}_{-0.02}$ \\
115 & 6.77948 & $-20.63^{+0.05}_{-0.04}$ & $9.14^{+0.26}_{-0.09}$ & $8.18^{+0.06}_{-0.06}$ & $1.27 \pm 0.06$ & $11.3^{+3.2}_{-2.0}$ & $8.15^{+0.10}_{-0.09}$ & $1.46^{+0.06}_{-0.01}$ \\
131 & 6.69836 & $-19.31^{+0.09}_{-0.09}$ & $8.18^{+0.57}_{-0.26}$ & $7.14^{+0.13}_{-0.11}$ & $2.42 \pm 0.39$ & $7.3^{+2.6}_{-1.1}$ & $3.11^{+0.10}_{-0.10}$ & $0.75^{+0.18}_{-0.09}$\,$^{\star}$ \\
\bottomrule
\end{tabular}
\caption{The OMEGA sub-sample presented in this work.
$M_{\rm UV}$ is the absolute magnitude at rest-frame 1500\,\AA, from a power law fitted to every ASTRODEEP-JWST band \citep{astrodeep} whose bandpass lies wholly redward of Lyman $\alpha$ at the source redshift and whose pivot falls in the rest-frame $1250$--$2600$\,\AA\ window, with the uncertainty from refitting the perturbed photometry; stellar masses are from \textsc{Prospector} SED fits described in Section~\ref{sec:sed_fitting}.
$12+\log(\mathrm{O/H})$ and $T_{\rm e}(\mathrm{O\,\textsc{iii}})$ are the direct-method values of Section~\ref{sec:chemical_abund}, the temperature quoted being the [O\,{\sc iii}]\,$\lambda4363$ route.
$\mathrm{O32} = [\mathrm{O\,\textsc{iii}}]\,\lambda5008\,/\,[\mathrm{O\,\textsc{ii}}]\,\lambda\lambda3727,3730$ and $R3 = [\mathrm{O\,\textsc{iii}}]\,\lambda5008\,/\,\mathrm{H}\beta$, both dereddened with each galaxy's own $A_V$. Limits on $O32$ are lower limits at $3\sigma$, [O\,{\sc ii}] being the undetected denominator; OMEGA-24 has no [O\,{\sc ii}] coverage.
$\mathrm{SFR}(\mathrm{H}\alpha)$ uses the \citet{clarke25} calibration on the H$\alpha$ flux, dereddened with each galaxy's own $A_V$ on the \citet{cardelli1989} law at $R_V = 3.1$, and so traces the last $\sim10$\,Myr.
Starred values come from H$\beta$ instead, through the Case~B ratio at that galaxy's own $T_{\rm e}$ and $n_{\rm e}$, where H$\alpha$ is unavailable: at $z = 8.28$ OMEGA-1 lies past the red end of G395M, and OMEGA-131 has no H$\alpha$ measurement.
All uncertainties are $16$--$84$th percentile ranges.}
\label{tab:sample}
\end{table*}

\subsection{Data reduction}
\label{sec:datared}

The data reduction process closely follows the one outlined in \citet{degraaff2024} and \citet{rai2026}, however we provide a summary here. Rather than adopt the reduced products released by each programme, we re-reduced all of the observations used in this work from the archive, such that the OMEGA spectra and the ancillary spectroscopy of Section~\ref{sec:ancillary} would be calibrated and extracted identically. 
Count-rate products for each programme were retrieved from the Mikulski Archive for Space Telescopes (MAST), having been processed through the \texttt{Detector1Pipeline} of the \texttt{jwst} pipeline. All subsequent processing used \texttt{msaexp}\footnote{\url{https://github.com/gbrammer/msaexp}} (v0.9.17; \citealt{msaexp}), which adopts the \texttt{jwst} (v1.16.1) pipeline procedures (e.g., wavelength calibration, flat-fielding, $1/f$ corrections, etc) and the \texttt{jwst\_1535.pmap} calibration files, providing consistent calibration and data reduction for each of the spectroscopic programs considered.
Moreover, we adopted the updated wavelength configuration of \texttt{msaexp}, which extends the nominal wavelength limits imposed by the STScI pipeline reference files. In particular, both G140M/F070LP and G140M/F100LP are extended redward to $\sim3 \mu$m, while G395M/F290LP is extended to $\sim5.5 \mu$m, effectively providing continuous wavelength coverage for our sources from $\sim0.7-5.5 \mu$m or $\sim1.0-5.5 \mu$m (using F070LP or F100LP, respectively). Second-order contamination is a concern only for G140M/F070LP, where light at $\lambda/2$ can pass the filter and be recorded at $\lambda$, affecting the spectrum redward of $\sim1.3\,\mu$m. For sources at $z>6$ in blank fields, such contamination is negligible, given the redshifted location of the Ly$\alpha$ break and absence of significant flux blueward.

For each source in a given MSA configuration, 2D spectral cutouts were extracted from the individual exposures and master background exposures determined for each nod position from the mean of the exposures in adjacent shutters. Background subtraction was then performed on the individual exposures using the mean background exposures, and 1D spectral extraction subsequently performed on the background-subtracted exposures using an optimal extraction procedure. In brief, the 1D spatial profile of each source was first modelled with an empirical point-spread-function (PSF) template, convolved to an intrinsic source profile of spatial width $\sigma$ and offset $y_{\lambda}$ relative to the centre of the spectral trace within the shutter. Both were left as free parameters and fit across all background-subtracted exposures simultaneously. Each 1D spectrum was then extracted optimally \citep{horne1986}, weighting each spectral element according to the fitted spatial profile and the inverse of its read-noise variance, before being reprojected onto a common wavelength grid for inverse-variance weighted combining into a final 1D spectrum.

\subsubsection{Slitloss Correction and Instrumental Resolution}

Slit loss corrections to account for the wavelength-dependent fraction of light falling outside the NIRSpec micro-shutter slitlet are computed using a morphology-based forward model approach with the \texttt{msafit} package \citep{degraaff2024}.
This package takes an intrinsic morphology model as input and computes the fractional slit losses at each wavelength based on the position of the source relative to the open microshutters.
For each source (see Table~\ref{tab:sample} and Figure~\ref{fig:cutouts}), we acquired available JWST/NIRCam imaging from the Dawn JWST Archive (DJA; \citealt{valentino2023}), using version 7.4 for the GOODS-N field.
We created $5^{\prime\prime}\times5^{\prime\prime}$ cutout images in each filter for which the source was covered.
We then masked out nearby sources using the segmentation map and fit an intrinsic morphology model in each filter using either a single (IDs 1, 22, 24, 115, 131) or double (IDs 6, 76, 77) 2D elliptical S\'{e}rsic profile, or a point source (ID 75), convolved with the NIRCam PSF in that filter.

If the integrated flux density in a filter had S/N$\ge$10, the resulting intrinsic morphology model was input into \texttt{msafit} along with the microshutter locations and orientations on-sky to model slit losses for the NIRSpec configurations that overlap in wavelength with each filter (e.g., the F290LP/G235M configuration spans a wavelength range that overlaps with the NIRCam filters that probe $\lambda>2.9~\mu$m).
For each filter, the fractional slit losses were saved only at those wavelengths covered by the filter.
The final slit loss correction function was fit with a fourth-order polynomial in wavelength to the computed slit losses from all available filters.
This approach accounts for wavelength-dependent effects due to both the increasing PSF size toward redder wavelengths and potentially different intrinsic morphologies at different rest-frame wavelengths since our imaging and spectroscopy spans from the far-UV to the red rest-frame optical.
The 1D science and error spectra were corrected by dividing by the wavelength-dependent fitted slit loss function.
The process above was repeated for each program that provides ancillary spectroscopy for the OMEGA targets since the microshutter positions and position angles differ between these programs, resulting in unique slit losses (see Figure~\ref{fig:cutouts}).

The \texttt{msafit} package also allows us to compute the effective spectral resolution for each source, accounting for the intrinsic morphology and location of the source relative to the shutters.
The NIRSpec spectral resolution provided in the JWST Documentation assume a uniformly illuminated slit.
For compact sources, the line spread function will be narrower than for the uniformly illuminated case leading to a higher effective spectral resolution.
For each NIRSpec configuration, we fit the spectral resolution estimates derived from the \texttt{msafit} model in all filters included above with a linear function in wavelength.
The spectral resolutions obtained in this way are typically $\approx1.5-1.7\times$ higher than the default resolutions for the OMEGA targets in this work.
These updated spectral resolutions were adopted when fitting the emission lines (Sec.~\ref{sec:linefitting}).

\subsubsection{Absolute Flux Calibration and Combining Spectra}

After slit loss corrections were applied, all spectroscopy used in this work was calibrated to the same absolute flux scale tied to the JADES DR5 photometric catalogue \citep{robertson26}.
For each NIRSpec configuration observed by each program used in this analysis, the 1D science spectra were passed through the transmission curves for each JWST or HST filter that the observed spectral wavelength range fully overlaps to calculate mock photometry.
The ratio of these mock photometric flux densities to the actual photometric measurements from imaging was computed, propagating uncertainties from both sets.
A scale factor was then computed by taking the inverse variance-weighted mean of the ratios for all filters in which both the mock and imaging photometry had S/N$\ge$3.
The 1D science and error spectra were then divided by this scale factor.
Note that this correction is a single factor applied to all wavelengths in a spectrum to tie the overall flux calibration to an absolute scale, which is distinct from the slit loss correction that provides a wavelength-dependent correction within each spectrum.
Both of these steps are required for robust line ratios across observations in more than one grating to compute robust physical properties from the combination of G140M and G395M spectra, and to properly combine data from multiple programs that observed the sources with different microshutter positions.

After this final flux calibration step, the agreement between flux densities in spectra taken in the same grating configuration by separate programs (e.g., PIDs 9214, 1914, 4762, and OMEGA all observed in G395M) was found to significantly improve.
We tested this agreement by comparing integrated fluxes across overlapping wavelength ranges covered by observations from different programs and line fluxes derived from Gaussian fitting in overlapping regions, and found a mean offset of 0.008~dex (1.8\%) and a dispersion of 0.07~dex (17\%).
This comparison demonstrates that our corrections have robustly accounted for differences introduced by the slightly differing shutter positions of different programs that observed OMEGA targets, such that we can robustly combine rest-UV information from archival G140M data with the OMEGA G395M spectra, and combine data from multiple programs in the same NIRSpec configuration when available to increase effective depth.

\begin{figure*}
    \centering
    \includegraphics[width=0.9\linewidth]{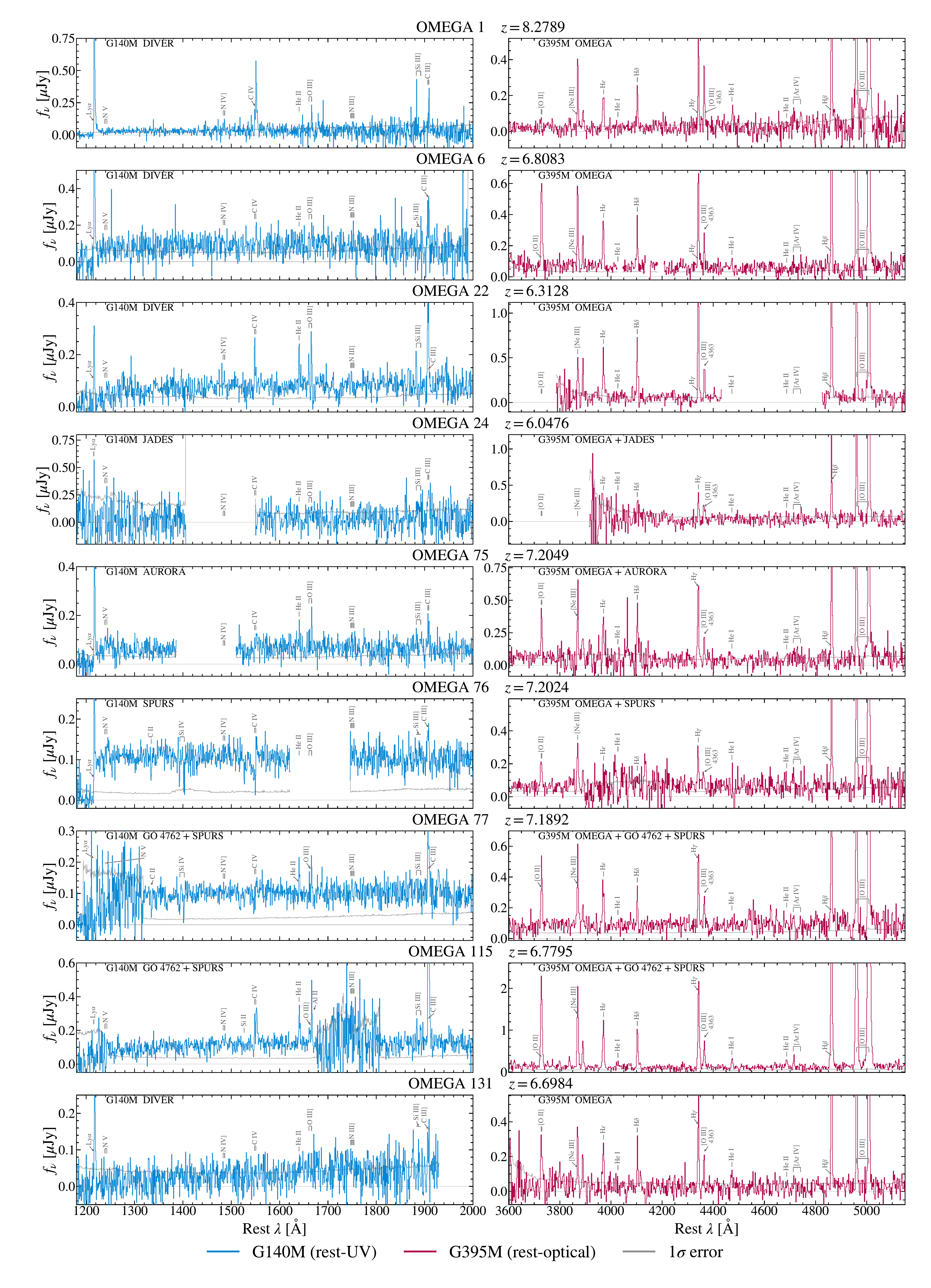}
    \caption{Rest-frame NIRSpec spectra of nine star-forming OMEGA galaxies in GOODS-N with ultra-deep rest-UV ancillary spectra. For each galaxy, the rest-UV (G140M) spectrum is shown above the rest-optical (G395M). Each plotted spectrum represents the inverse-variance mean of all available G140M (blue) or G395M (red) spectra. The $1\sigma$ error spectrum is plotted in each panel in grey, while the wavelengths of emission lines of interest are indicated with grey text and lines.}
    \label{fig:spectra}
\end{figure*}

The final science spectra analysed in this work, presented in Figure~\ref{fig:spectra}, were constructed by inverse-variance mean combining the available spectra in each grating of interest (G140M or G395M) from OMEGA and all of the programs listed in Sec.~\ref{sec:ancillary}.
The final error spectra were computed using the standard error on an inverse-variance weighted mean based on the individual error spectra, and are found to be well matched to the variance of flux densities in blank regions of the spectra.
Table~\ref{tab:programs} provides the programs and total integration time contributing to the G140M and G395M spectrum of each source in this analysis.

\begin{table}
\centering
\renewcommand{\arraystretch}{1.3}
\begin{tabular}{lcc|cc}
\toprule
OMEGA & \multicolumn{2}{c|}{G140M} & \multicolumn{2}{c}{G395M} \\
\cline{2-3}\cline{4-5}
MSAID & $t_\mathrm{int,tot}$ (h) & PIDs & $t_\mathrm{int,tot}$ (h) & PIDs \\
\midrule
1 & 20.4 & 8018 & 13.1 & 7729 \\
6 & 20.4 & 8018 & 13.1 & 7729 \\
22 & 20.4 & 8018 & 13.1 & 7729 \\
24 & 0.9 & 1181 & 14.0 & 7729, 1181 \\
75 & 12.2 & 1914 & 17.3 & 7729, 1914 \\
76 & 29.2 & 9214 & 16.0 & 7729, 9214 \\
77 & 31.2 & 9214, 4762 & 18.1 & 7729, 9214, 4762 \\
115 & 31.2 & 9214, 4762 & 18.1 & 7729, 9214, 4762 \\
131 & 20.4 & 8018 & 13.1 & 7729 \\
\bottomrule
\end{tabular}
\caption{Total on-source integration times and contributing programme IDs for the final G140M and G395M spectra analysed in this study.}
\label{tab:programs}
\end{table}

\section{Methods}
\label{sec:methods}

\subsection{Spectroscopic Measurements}
\label{sec:linefitting}

Emission-line fluxes and equivalent widths were measured by fitting Gaussian line profiles to the 1D science spectra.
The set of strong lines H$\beta$, [O\,{\sc iii}]\,$\lambda\lambda4960,5008$ \AA, and H$\alpha$ were first fit with single Gaussians and a linear continuum to determine the systemic redshift and an intrinsic velocity FWHM after correcting for the instrumental resolution, with the adopted values taken as the inverse variance-weighted mean of the values inferred for this set of lines.
Emission lines were then fit with Gaussian profiles, where the centroid is bound within $\pm50$~km/s of the systemic redshift and the FWHM is restricted to be within 20\% of the width implied by the intrinsic velocity FWHM computed above and above the instrumental resolution at that wavelength.
Closely-spaced lines with $\Delta\lambda/\lambda<0.01$ were fit simultaneously, and lines in doublets from the same ion (e.g., C\,{\sc iii}]\,$\lambda\lambda1907,1909$ \AA, [O\,{\sc ii}]\,$\lambda\lambda3727,3730$ \AA) had their line widths and redshifts tied to one another.
The continuum is taken to be the combined stellar and nebular continuum drawn from the SED fitting described below (Sec.~\ref{sec:sed_fitting}), with a multiplicative scale factor that allows fine-tuning of the continuum level around each set of emission lines and allows for the propagation uncertainty in the continuum measurement into line flux and equivalent width uncertainties.
Uncertainties on all measured emission-line properties were taken to be half of the 16th-84th percentile width of the parameter distributions acquired by perturbing the spectrum by the error spectrum 1000 times and re-measuring all emission lines in each realisation.

\subsection{Spectral Energy Distribution Fitting}
\label{sec:sed_fitting}

The target spectral energy distributions (SEDs) were modeled using the {\sc Prospector} SED fitting code \citep{2021ApJS..254...22J}, based on photometry from the JADES DR5 photometry \citep{robertson26}. The photometric catalogue was constructed based on imaging taken with the JWST/NIRCam, HST/ACS, and HST/WFC3 instruments in 27 filters: F070W, F090W, F115W, F150W, F162M, F182M, F200W, F210M, F277W, F300M, F335M, F356W, F410M, F430M, F444W, F460M, F480M, F435W, F606W, F775W, F814W, F850LP, F105W, F125W, F140W, and F160W. In addition to the JADES program, imaging from the FRESCO \citep{2023MNRAS.525.2864O}, BEACON \citep{2025ApJ...983..152M}, and PANORAMIC \citep{2025ApJ...979..140W} surveys were incorporated into the photometric measurements presented in the catalogue for the GOODS-N objects. From the JADES DR5 photometric catalogue, we utilised measurements from the \texttt{KRON\_CONV} extension, which consists of photometry measured within Kron apertures on imaging that has been convolved to match the F444W point-spread function.

To fit the photometry, we used {\sc Prospector} with the Flexible Stellar Population Synthesis package \citep[FSPS;][]{2009ApJ...699..486C, 2010ApJ...712..833C}, MILES stellar spectral library \citep{2006MNRAS.371..703S} and MIST isochrones \citep{2011ApJS..192....3P,2013ApJS..208....4P,2015ApJS..220...15P,2016ApJS..222....8D,2016ApJ...823..102C}. We adopted an SMC dust law \citep{2003ApJ...594..279G}, a \citet{2003PASP..115..763C} initial-mass function (IMF), and a non-parametric star-formation history (SFH). We divided this non-parametric SFH into eight time bins, utilising a Student-t distribution as a continuity prior \citep[e.g.,][]{tacchella22}. The two most recent bins spanned the previous 3 Myr and 10 Myr, respectively, and the remaining six bins were distributed in even logarithmically spaced time intervals, reaching a maximum lookback redshift of $z=20$. We additionally allowed the stellar metallicity to vary as a free parameter.

The SEDs were fit using models that incorporated both stellar and nebular emission. The nebular component of the SED models was drawn from a pre-computed grid of {\sc Cloudy} photoionisation models in which the gas-phase metallicity is tied to the stellar value \citep{2013RMxAA..49..137F,2017ApJ...840...44B}. The resulting best-fit stellar and nebular continuum SED components from these fits were then used to model the continuum when fitting emission lines in the flux-calibrated grating spectra.

\subsection{Derivation of Accurate Chemical Abundances}
\label{sec:chemical_abund}

\subsubsection{Stratified ISM Framework}
\label{sec:stratified}

Recent studies have shown the ISM gas of high-redshift galaxies to be multiphase and stratified \citep[e.g.][]{harikane2025, takechi2026, moreschini2026, mendezdelgado2023, martinez2025}, with over-simplistic single-zone frameworks significantly biasing derived properties. As such, we first characterise the gas via the multi-zone scheme of \citet{martinez2025}, following the implementation of \citet{rai2026}, to which we refer the reader for details.

In brief, each ion is assigned to the low-, intermediate- or high-ionisation zone in which it forms, whether or not it is detected in the spectrum, and each zone carries its own electron temperature and density. Densities are derived from emission line doublet ratios (e.g., \niv]\,$\lambda\lambda1483,1486$ \AA, \ciii]\,$\lambda\lambda1907,1909$ \AA\ or [\sii]\,$\lambda\lambda6718,6732$ \AA) where available, or from redshift-dependent trends when necessary \citep{abdurrouf2024,martinez2025}. High-ionisation zone temperatures ($T_{e,\rm high}$) are measured using the [\oiii]$\lambda$4363 \AA\ line from the OMEGA spectra, while the low- and intermediate-ionisation zone temperatures are inferred from $T_{e,\rm high}$ using the relations of \citet{garnett1992}. All emission line fluxes were first de-reddened using the \citet{cardelli1989} extinction and measured Balmer decrements, prior to their in $T_{e}$ or $n_{e}$ inferences.

The ionisation parameter $U$, defined as the ratio of ionising photon density to hydrogen density, regulates the ionisation state of the gas and therefore the ionisation correction factors (ICF) required to convert observed ionic ratios into chemical abundance ratios \citep{martinez2025}. We adopt $\log U$ from the \textsc{Prospector} fits of Section~\ref{sec:sed_fitting} to determine the ICF required to convert the ionic ratios of our detected emission lines into N/O and C/O abundance ratios. The electron temperatures, electron densities, and ionisation parameters for each galaxy and ionisation zone are listed in Table~\ref{tab:conditions}.

\subsubsection{C-N-O Abundances}

As mentioned in the previous section, all line fluxes are first corrected for dust attenuation using the \citet{cardelli1989} extinction law with $R_V = 3.1$, and the multiple available Balmer decrements enabled by our medium-resolution spectroscopy. The choice of attenuation curve has little bearing on our results: four of the nine galaxies in our sample are consistent with no attenuation and the largest colour excess is $E(B-V) = 0.13$, so adopting the nebular curve of \citet{reddy2020} in the optical, or a starburst \citep{calzetti2000} or LMC curve \citep{gordon2003} across the rest-UV, shifts our line ratios by less than $0.05$\,dex, within their measurement uncertainties.

Gold-standard O abundances using electron temperatures from [O\,{\sc iii}]\,$\lambda4363$ \AA\ in individual galaxies represents the primary goal of the OMEGA survey, and provides a key anchoring for precise N/O and C/O abundance ratios. To this end, the detection of the emission line in each of our nine galaxies in our sample (by construction), highlights the achievability of such an endeavour without recourse to stacking, and also allows for comparison to rest-UV auroral line measurements. Where O\,{\sc iii}]\,$\lambda1666$ is also detected in the rest-UV spectra (OMEGA-1, 22, 75, 77 and 115), we compare the inferred electron temperatures and metallicities from both auroral lines.

\begin{figure*}
    \centering
    \includegraphics[width=0.8\linewidth]{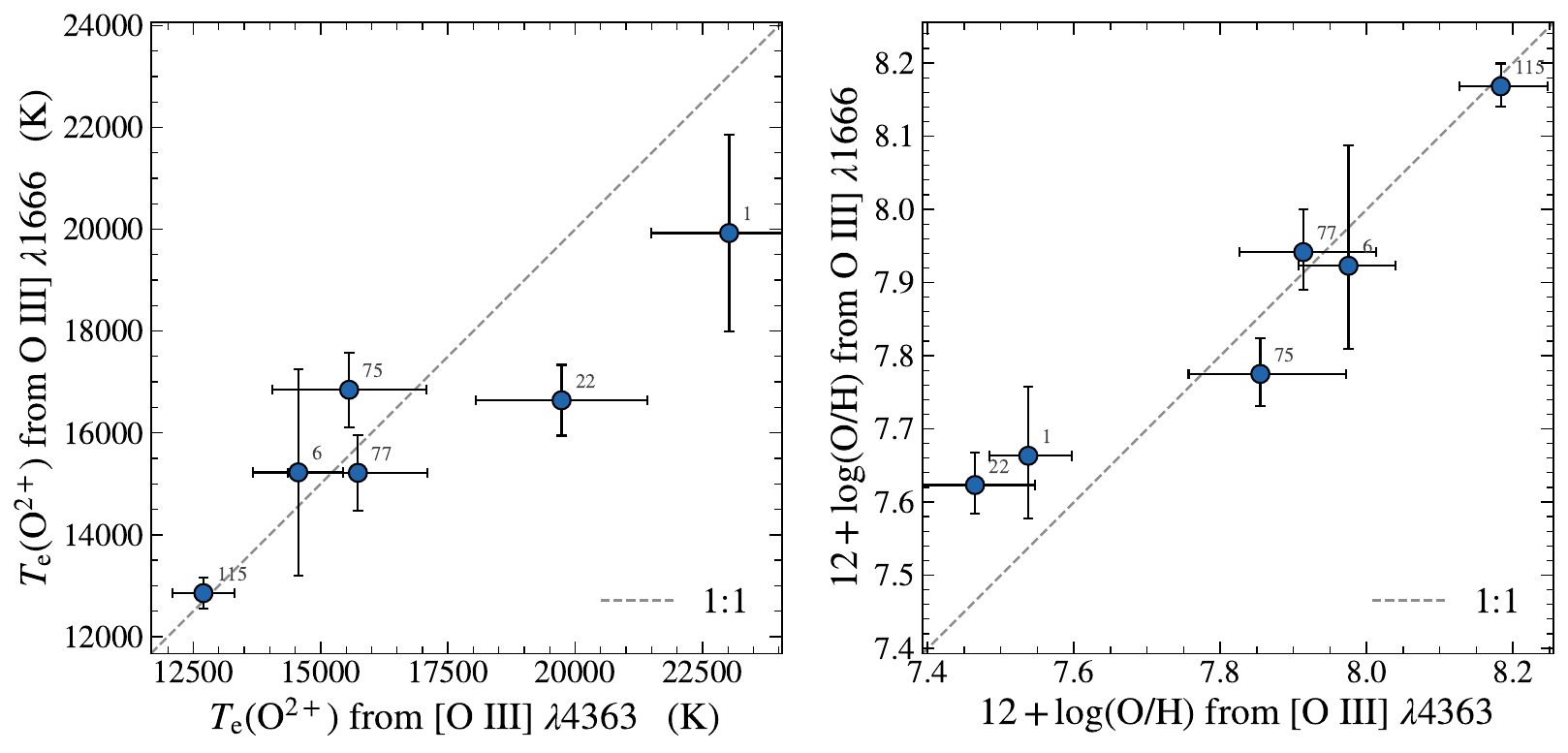}
    \caption{\textsc{Left}:  Electron temperature measured from the two auroral lines of O$^{2+}$, ultraviolet O\,{\sc iii}]\,$\lambda1666$ and optical [O\,{\sc iii}]\,$\lambda4363$. 
    \textsc{Right}: Direct-$T_{\rm e}$ $12+\log(\mathrm{O/H})$ that follows from each auroral line.  The six galaxies with a measurement of O\,{\sc iii}]\,$\lambda1666$ are plotted, with the dashed line marking the 1:1 relation. The two methods are broadly in agreement, with every galaxy sitting within $2\sigma$ of the 1:1 relation in both panels.}
    \label{fig:te_consistency}
\end{figure*}

The electron temperatures are computed using the collision strengths of \citet{tayal2017} and shown in the left panel of Figure~\ref{fig:te_consistency}, with excellent agreement within $2\sigma$. O/H abundance is then derived from the sum of its two ionic abundances, $\mathrm{O/H} = \mathrm{O^{+}/H^{+}} + \mathrm{O^{2+}/H^{+}}$, with no ionisation correction. O$^{2+}$ is calculated using [O\,{\sc iii}]\,$\lambda\lambda4960,5008$ with $T_{\rm e}^{\rm high}$, and O$^{+}$ from [O\,{\sc ii}]\,$\lambda\lambda3727,3730$ with $T_{\rm e}^{\rm low}$ (each adopting the electron density characteristic of their ionisation zone). [O\,{\sc ii}] is detected in all but three of our sources (OMEGA-1, -22, and -24, the latter due to the NIRSpec detector gap), however at the ionisation parameters typical of our sample contributes very little to the total oxygen budget: O$^{+}$ accounts for only $0.6-11.0$ per cent of O/H across the full sample. As such, neglecting the O$^{+}$ term altogether would therefore change $12+\log(\mathrm{O/H})$ by at most $0.05$\,dex. As shown in the right panel of Figure~\ref{fig:te_consistency}, the metallicities calculated using electron temperatures from either O\,{\sc iii}]\,$\lambda1666$ \AA\ or [O\,{\sc iii}]\,$\lambda4363$ \AA\ are in excellent agreement.

N/O is measured as N$^{+}$/O$^{+}$ from [N\,{\sc ii}]\,$\lambda6585$ \AA\ and the [O\,{\sc ii}] doublet, corrected to total N/O with the \citet{martinez2025} ICF for those transitions. [N\,{\sc ii}]\,$\lambda6585$ is detected in two galaxies, OMEGA-6 and 115, while N\,{\sc iv}]\,$\lambda\lambda1483,1486$ \AA\ is detected in only one source (OMEGA-1). For this latter object, we determine N/O from N$^{3+}$/O$^{2+}$ with the corresponding \citet{martinez2025} ICF, which, at $1.68$, is significantly larger than the $0.82$ to $0.95$ ICFs required for N$^{+}$/O$^{+}$ corrections in the [\nii]-detected objects. 

C/O is determined from C$^{2+}$/O$^{2+}$ using the ratio of C\,{\sc iii}]\,$\lambda\lambda1907,1909$ \AA\ to either [O\,{\sc iii}]\,$\lambda\lambda4960,5008$ \AA\ or, where detected, O\,{\sc iii}]\,$\lambda1666$ (each requiring an appropriate ICF to obtain total C/O). The two C/O measurements (based on rest-UV or rest-optical O\,{\sc iii}] lines) agree within $0.16$\,dex in the six galaxies for which both measurements are available.  Due to the proximity of the [\ciii] and O\,{\sc iii}]\,$\lambda1666$ \AA\ emission lines, we adopt these as our fiducial C/O diagnostics when available, to avoid any remaining systematic or redenning effects which may not be fully accounted for. In the case of O\,{\sc iii}]\,$\lambda1666$ \AA\ non-detections, we instead utilise [\oiii]$\lambda4363$ \AA.

In cases where emission lines remain undetected, we calculate $3\sigma$ upper limits using the uncertainties from the line measurements described in Section~\ref{sec:linefitting}. We opt for line fitting uncertainties instead of fixed-width matched-filter estimates, which account only for the uncertainty on the line amplitude and not for the width, a contribution that persists even at high signal-to-noise. This limit is then propagated through all relevant abundance calculations.

\section{A robust sample of emission-line galaxies at \texorpdfstring{$\bm{z \geq 6}$}{z >= 6}}
\label{sec:demographics}

While our sample represents only a small subset of the total OMEGA survey, the richness of the ultra-deep rest-UV-to-optical spectroscopy of each of our sources enables us to evaluate the detection rate of key emission/absorption features, robustly assess their star-forming versus AGN nature, and place their properties in context relative to the global population of $z\sim6-8$ galaxies.

\subsection{Spectral features revealed from ultra-deep observations}

As described above, the sources in our sample were selected based on a minimum of [\oiii]$\lambda$4363 \AA\ and \ciii]$\lambda\lambda$1907,1909 \AA\ detections. Unsurprisingly given the depth of these observations, we detect at $\geq3\sigma$ the lines measured routinely in $z>6$ galaxies: \hbeta, H$\gamma$ and [\oiii]$\lambda\lambda$4960,5008 \AA\ in all nine galaxies, the H8$+$\hei\,$\lambda$3889 \AA\ blend in seven galaxies, \hei\,$\lambda$5876 \AA\ and [\neiii]$\lambda$3870 \AA\ in seven galaxies, and [\oii]$\lambda\lambda$3727,3730 \AA\ in six of the eight galaxies in which the line is covered.
As seen in Figure~\ref{fig:spectra}, the same depth also affords a far richer suite of emission lines, absorption profiles, and stellar wind features that are beyond the typical depths of archival NIRSpec $R\sim1000$ spectra or the resolution of NIRSpec prism spectroscopy.

\lya\ is detected in six of the nine sources (OMEGA-1, 6, 22, 75, 76 and 131), suggesting that these sources sit within ionised bubbles large enough for the line to redshift out of resonance before encountering intervening neutral hydrogen along the line of sight of the intergalactic medium \citep{dijkstra2014,mason2018}.

\civ\,$\lambda\lambda$1548,1551 \AA\ appears in emission in three of the eight galaxies with coverage (OMEGA-1, 22 and 115) and OMEGA-1 is the only galaxy of the nine in which \civ\ exceeds \ciii]\ in strength (with rest frame EW $27.2$\,\AA\ versus $18.2$\,\AA, respectively). Strong \civ\ has been associated with intense bursts of star formation \citep[e.g.,][]{rai2026,robertsborsani2026}, and all three emitters sit above the star-forming main sequence at $z\sim6-8$ ($+0.81$, $+0.80$ and $+0.46$\,dex for OMEGA-1, 22 and 115, respectively, see Section~\ref{subsec:sfms}), as shown in Figure~\ref{fig:sfms}.

\niv]\,$\lambda\lambda$1483,1486 \AA\ is detected at $>3\sigma$ in only one source with appropriate coverage, OMEGA-1. This source also shows the strongest \civ\ of the entire sample, as described above. Both lines require a hard ionising field and are frequently detected in composite spectra, the likes of which display an apparent trend of \niv\ detections in the strongest \civ\ emitters \citep{rai2026,robertsborsani2026}.

\heii\,$\lambda$1640 \AA\ is also detected in four of the eight galaxies with coverage (OMEGA-22, 75, 77 and 115), and \heii\,$\lambda$4686 \AA\ in one (OMEGA-115 with $S/N=5.0$). Interestingly, in OMEGA-115, the line fits to both helium lines are $26\pm7$ and $27\pm13$ per cent broader than the fit to [\oiii]$\lambda$5008 \AA, with $351\pm20$\,\kms\ and $354\pm35$\,\kms\ versus $278\pm1$\,\kms, similar to possible broadening signatures associated with Wolf-Rayet and/or VMS spectra at comparable redshifts \citep{berg2025,marqueschaves2026}. Such broadening alone does not provide conclusive evidence for exotic stellar origins, while the routine detections of the line in $z\gtrsim6$ composite spectra highlight an important nebular component in early galaxy spectra \citep{rai2026,umeda2026}. Nonetheless, combined with \heii\,$\lambda$4686/\hbeta\ ratios that exceed the levels of standard BPASS stellar population models (0.041$\pm$0.008 versus 0.003, a factor of $>13\times$ increase), such broadening in individual sources could plausibly be indicative of Wolf-Rayet or VMS contributions, particularly if paired with clear ``blue bump'' emission line signatures \citep{berg2025}.   

The ultra-deep nature of the rest-UV spectroscopy also reveals prominent absorption features from stellar winds and the ISM. The clearest wind features are associated with OMEGA-76, the brightest source in our sample, where both \nv\ and \civ\ lines are seen as a combination of blueshifted absorption and redshifted emission characteristic of P-Cygni profile. Comparable P-Cygni profiles have recently been presented by \citet{marqueschaves2026} in two luminous $z>8$ examples, which they attributed to Very Massive Stars. ISM absorption, on the other hand, is observed in both low- and high-ionisation transitions at various significance. The low-ionisation gas is traced by O\,{\scriptsize I}\,$\lambda$1302 \AA, Si\,{\scriptsize II}\,$\lambda\lambda$1304,1526 \AA, \cii\,$\lambda$1334 \AA, Fe\,{\scriptsize II}\,$\lambda$1608 \AA\ and Al\,{\scriptsize II}\,$\lambda$1670 \AA, while the high-ionisation gas is traced by Si\,{\scriptsize IV}\,$\lambda\lambda$1394,1403 \AA. We model (covering fractions and velocity offset) the two gas phases separately (incorporating all lines of a given phase), according to the procedure outlined by \citet{vasan2026} and weight them according to their inverse-variance arrays. High ionisation Si\,{\scriptsize IV}\,$\lambda\lambda$1394,1403 \AA\ absorption is seen in two source (OMEGA-76 and 77), with net blueshifts of $-181\pm51$\,\kms\ and $-192\pm29$\,\kms\ relative to systemic velocities, suggestive of outflowing gas. Low ionisation \cii\,$\lambda$1334 \AA\ is instead seen in three of our nine galaxies (OMEGA-76, 77 and 115) but characterised by milder velocity shifts close to systemic velocities of $-74\pm53$\,\kms\ to $+115\pm64$\,\kms. Such differences are similar to those reported in \citet{vasan2026} for a number of individual $z>5$ sources at high-redshift, and the stacked $z>6$ spectra analysed by \citet{rai2026}, suggestive of multi-phase outflows.

\subsection{UV Emission-line diagnostics indicating an absence of AGN}
\label{sec:uvbpt}

Although our sample of $z>6$ sources were selected a priori to not harbour any clear AGN signatures (e.g., broad Balmer lines or LRD-like features), complete purity cannot be guaranteed and it is therefore instructive to verify their positions on various UV emission-line diagnostic diagrams, relative to the loci of star-forming sources and narrow-line AGN. This is particularly interesting at the redshifts probed by our sample, where recent JWST studies have demonstrated the ineffectiveness of traditional (optical) BPT diagnostics in discerning the hard radiation fields of low-metallicity galaxies from narrow-line AGN \citep[e.g.,][]{jadesagn2}.

\begin{figure*}
    \centering
    \includegraphics[width=1\linewidth]{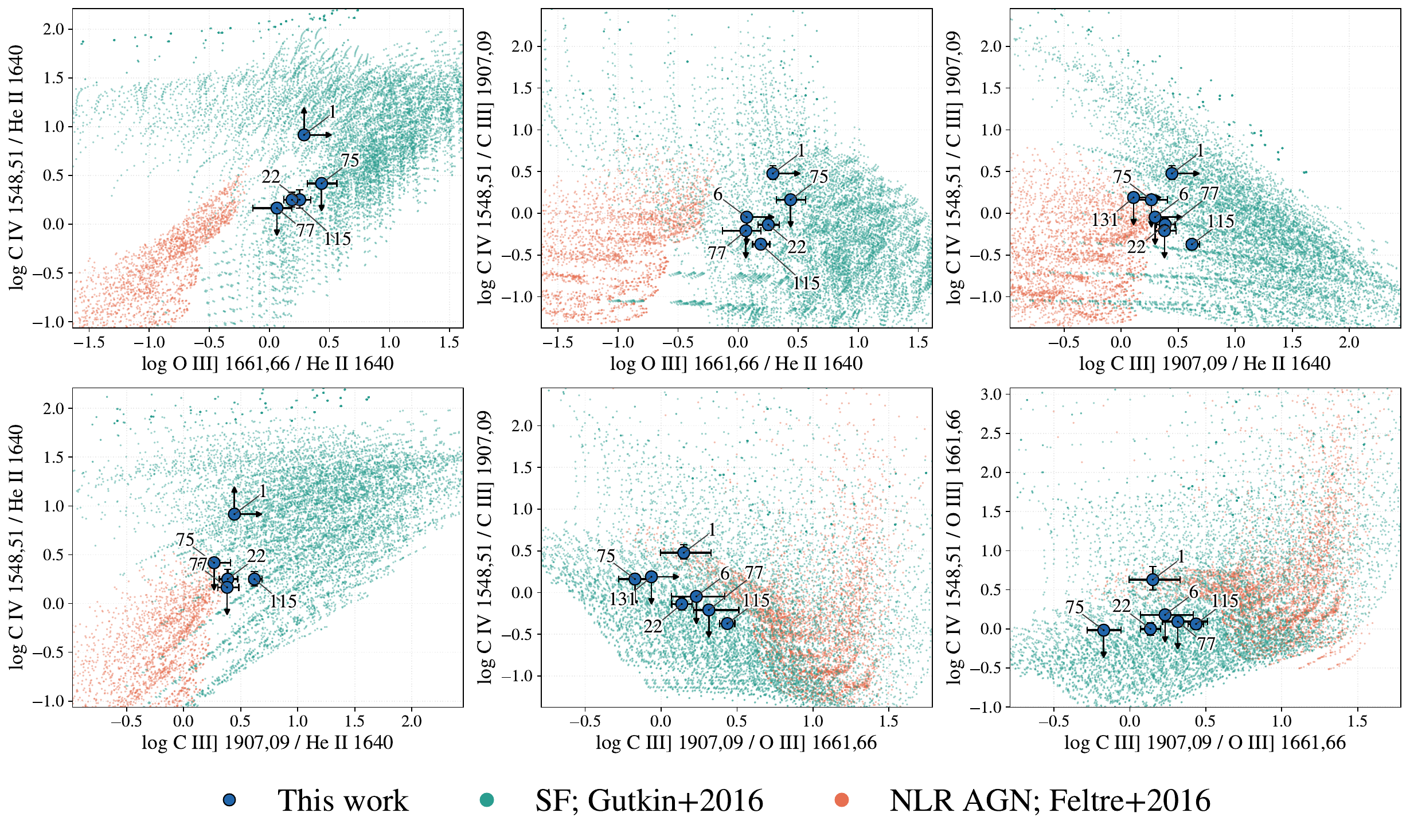}
    \caption{Rest-UV diagnostic planes of dust-corrected line ratios for the nine star-forming OMEGA galaxies in GOODS-N, with $3\sigma$ limits where a line is not detected. The star-forming grids of \citet{gutkin2016} and the narrow-line AGN grid of \citet{feltre2016} are shown in teal and red, respectively.}
    \label{fig:uv_diagnostics}
\end{figure*}

In Figure~\ref{fig:uv_diagnostics} we present six rest-UV line ratio diagrams based on a combination of high-ionisation \civ, \ciii], \oiii], and \heii, along with the measurements for each of our sources. For reference, and following \citet{topping2025}, we also showcase the star-forming models of \citet{gutkin2016} and the narrow-line AGN models of \citet{feltre2016}. Although there is some minor overlap in models (depending on the chosen diagnostic), every source in our sample -- include those limited by lower/upper limits -- falls within the locus of star-forming sources. While some diagnostics show this more clearly than others (e.g., \civ/\heii or \civ/\ciii] vs. \oiii]/\heii, compared to \civ/\heii or \civ/\ciii] vs. \ciii]/\heii), this observation holds for every diagnostic considered; in other words, there are no sources that are considered unambiguously star-forming with one BPT diagnostic but AGN in another, as has been seen at moderately higher redshifts \citep{castellano2024,napolitano25}. The closest possible exception to this rule is OMEGA-1, which is the clearest star-forming candidate in \heii-based diagnostics (due to non-detection of the line), but lies close to the edge of the AGN locus in examples where \civ, \ciii], and \oiii] are detected and utilised.

Nonetheless, the sample as a whole shows no compelling evidence for AGN activity based on their UV line ratios, and the test highlights both the discerning power of these emission-line diagnostics as well as the necessity for ultra-deep spectroscopy, where higher ionisation lines can be confidently detected or strong upper limits placed.

\subsection{Bursty galaxies above the Star-Forming Main Sequence}
\label{subsec:sfms}
Our sample of nine galaxies spans stellar masses $\log(M_\star/\mathrm{M}_\odot) \approx 7.9$--$9.1$ (median $\log(M_\star/\mathrm{M}_\odot)=8.24$) and absolute UV magnitudes $M_{\rm UV} \approx -18.7$ to $-21.1$ (median $M_{\rm UV}=-20.17$), and as such can be considered part of the ``moderately-bright'' population. In order to place our sample into further context of the global population of galaxies within the first billion years, we evaluate their positions relative to the star-forming main sequence (SFMS) at high redshift, whereby galaxies display a strong correlation between their measured star formation rate (SFR) and stellar mass \citep{noekse07}. While there is some debate regarding the exact formation epoch of the SFMS at high redshift, the relation appears to be well in place by $z\sim8$ \citep{nakajima2023,simmonds2025,cole25,ciesla2024,clarke24}, with scatter that reflects the increasingly bursty star formation histories of galaxies at these epochs \citep{ciesla24bursty}.

\begin{figure}
    \centering
    \includegraphics[width=1\linewidth]{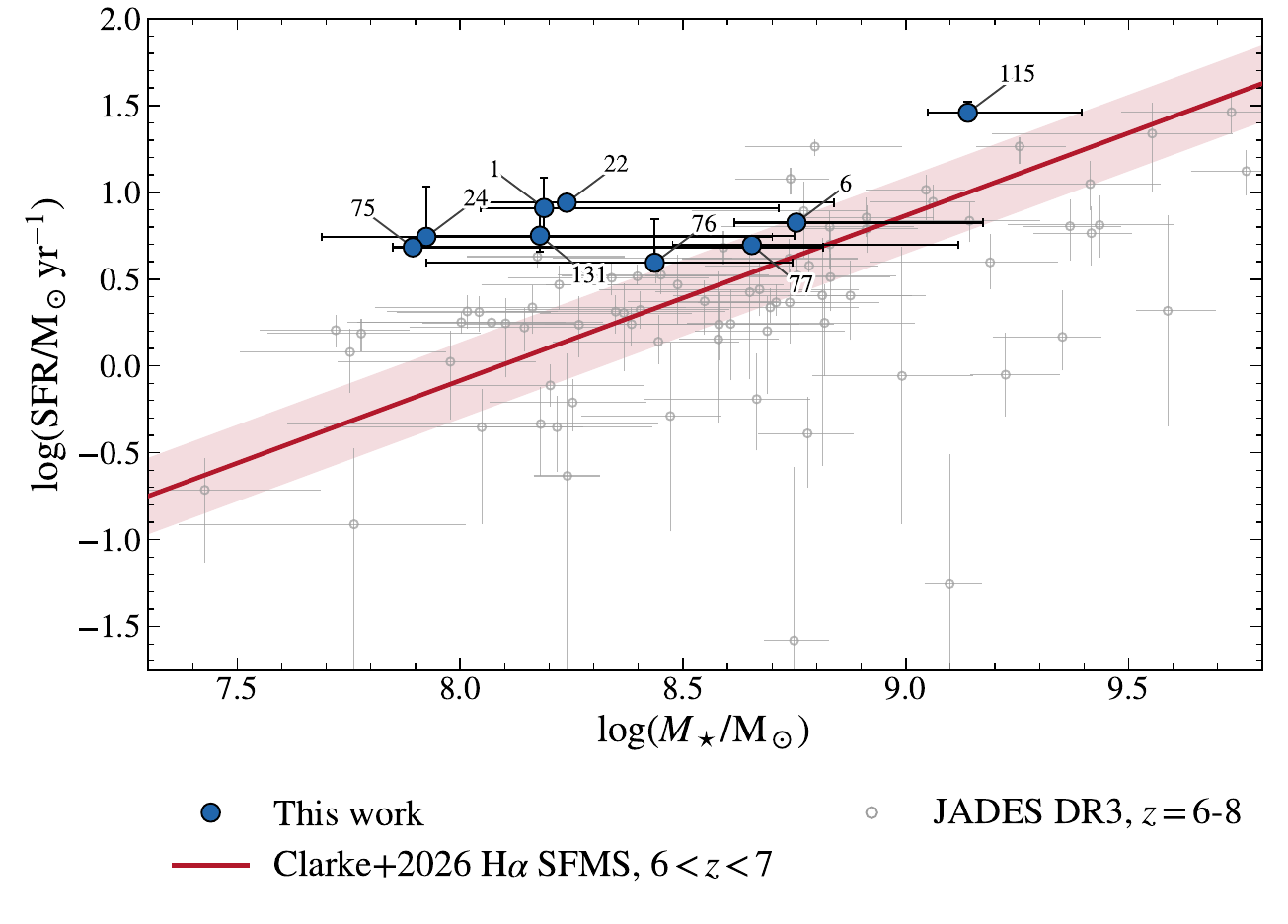}
    \caption{Star-formation rate vs. stellar mass for the nine OMEGA galaxies in our sample. Stellar masses are from the \textsc{Prospector} fits of Section~\ref{sec:sed_fitting}.  H$\alpha$ is unmeasured in two galaxies (OMEGA-1 and 131). For these galaxies, star-formation rates are based on H$\beta$, with the assumption of the   Case-B H$\alpha$/H$\beta$ ratio appropriate for each galaxy's specific  $T_{\rm e}$ and $n_{\rm e}$. The red line is the H$\alpha$ star-forming main sequence in the $6 < z < 7$ region calibrated by \citet{clarke25}. This calibration, however, is not fit below $10^{8.5}\,\mathrm{M}_\odot$. Grey points are the $78$ JADES DR3 galaxies \citep{deugenio25} at $z = 6$--$8$ used as a comparison sample, where star-formation rates are the $10$\,Myr averages of their own SED fits.}
    \label{fig:sfms}
\end{figure}

In Figure~\ref{fig:sfms}, we plot the OMEGA sources on the SFMS, comparing to the $6<z<7$ SFMS presented in \citet{clarke25}. The background gray points show the distribution objects at $6<z<8$ from JADES DR3 \citep{deugenio25,clarke25}, where the SFR measurements are determined from the {\sc Prospector}-based SED fits averaged over the most recent 10 Myr. Although the stellar mass uncertainties are large (up to $\sim0.5$ dex), this subset of the full OMEGA sample lies above or on the upper end of the SFMS. This offset is unsurprising, given that the targets were selected to have bright emission lines, enabling the detection of their faint auroral-line features. The offset is particularly pronounced at lower masses of $\log(M_\star/\mathrm{M}_\odot)<8.5$, where galaxies at high redshift are expected to represent the burstiest systems \citep{endsley2025}. At the lower stellar-mass-half probed by our sample ($7.9\lesssim\log(M_\star/\mathrm{M}_\odot)\lesssim8.5$), the OMEGA sources sit between $0.3$ and $0.9$\,dex above the SFMS parametrization of \citet{clarke25}, with a median offset of $0.8$\,dex. 
Such differences decrease at higher stellar masses ($\log(M_\star/\mathrm{M}_\odot)\gtrsim8.5$, where the \citealt{clarke25} sample is complete), where the median offset decreases to $+0.2$\,dex above the SFMS, comparable to the $0.22$\,dex intrinsic scatter \citet{clarke25} measure for the relation in this redshift bin.

\section{Accurate Metallicities and Elemental Abundances in the First Billion Years}
\label{sec:metals}

The primary goal of this analysis (and OMEGA as a whole) is to demonstrate the power of ultra-deep, $R\sim1000$ spectroscopy in delivering precise chemical abundances for individual galaxies, 
as opposed to the stacking of shallower data. In this respect, [\oiii]\,$\lambda4363$ \AA\ is detected in all nine of the galaxies presented here, with S/N between $3.3$ and $10.8$. As outlined above, these detections enable $T_{\rm e}$ determinations spanning $1.27-2.42\times10^{4}$\,K. Applying the framework outlined in Section~\ref{sec:chemical_abund}, we present here the metallicities and elemental abundance ratios for our sample, listed in full in Table~\ref{tab:conditions}.

\subsection{Oxygen abundances and the mass-metallicity relation}

\begin{figure}
    \centering
    \includegraphics[width=1\linewidth]{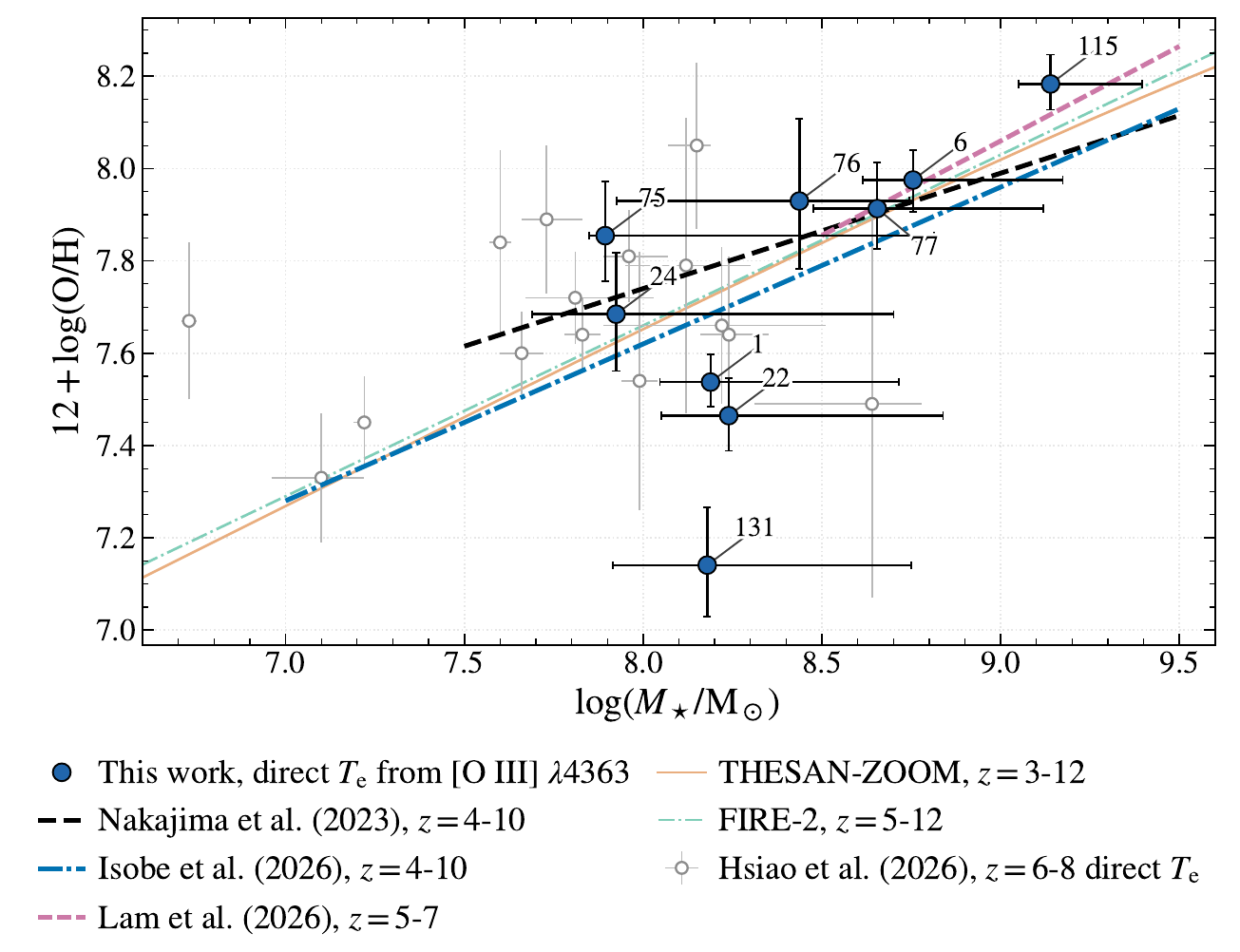}
    \caption{Mass-metallicity relation (MZR) for the nine galaxies in our sample. Stellar masses from the \textsc{Prospector} fits of Section~\ref{sec:sed_fitting} and oxygen abundances from the direct method, every one of them resting on a detection of [\oiii]\,$\lambda4363$. Open grey circles are the $z = 6$--$8$ direct-$T_{\rm e}$ compilation of \citet{hsiao2026b}. Heavy curves are observed relations, each drawn only over the stellar mass interval its authors fit: the $z = 4$--$10$ relation of \citet{nakajima2023} in black, the $z = 4$--$10$ relation of \citet{isobe2026} in blue, and the $z = 5$--$7$ relation of \citet{lam2026} in pink. All three observed relations rest on strong-line calibrations: \citet{nakajima2023} measure 8 of their 135 galaxies with the direct method and the remainder from $R23$ and $R3$, \citet{isobe2026} fit the median of individual strong-line metallicities in bins of stellar mass, and \citet{lam2026} measure theirs from stacked spectra. Every point in this work instead carries its own electron temperature. The THESAN-ZOOM \citep{mcclymont2025} and FIRE-2 \citep{marszewski2024} predictions are plotted in the background in orange and green respectively, evaluated at the sample median redshift of $z = 6.81$.}
    \label{fig:mzr}
\end{figure}

We measure a range of direct-method metallicities spanning $12+\log(\mathrm{O/H}) = 7.14$--$8.18$, with a median of $7.85$. Figure~\ref{fig:mzr} presents these on the mass-metallicity plane (MZR), along with the observed high-redshift relations of \citet{nakajima2023}, \citet{isobe2026} and \citet{lam2026}, as well as the predictions from the THESAN-ZOOM \citep{mcclymont2025} and FIRE-2 \citep{marszewski2024} simulations. Six of our nine galaxies sit within $0.2$\,dex of the \citet{nakajima2023} relation, while OMEGA-1, 22, and 131 lie $0.25$, $0.34$ and $0.64$\,dex below that relation. 

All three sit within $0.06$\, dex of the same stellar mass, $\log(M_\star/\mathrm{M}_\odot) \simeq 8.2$, highlighting a diversity of oxygen abundances within our sample compared to other samples at comparable redshifts (c.f. with \citealt{morishita24}). These three galaxies share a high electron temperature, $1.97-2.42\times10^{4}$\,K, which in the direct method propagates to a lower metallicity.
These galaxies have among the larger positive offsets in the sample, in terms of sSFR relative to the SFMS. According to the galaxy Fundamental Metallicity Relation \citep{ellison2008,mannucci2010,curti2020,curti2024}, a positive offset in sSFR corresponds to a negative offset in gas-phase metallicity, which could be used to understand the lower-than-average metallicities for these galaxies. At the same time, OMEGA-24 and 75 have comparable positive offsets relative to the SFMS, and yet their metallicities are entirely constant with the MZR \citep{nakajima2023}. A full analysis of the normalisation, slope, and scatter of the $z>6$ MZR will be presented in future OMEGA works that include the full sample.

\subsection{High-ionisation line strengths vs. metallicity}

\begin{figure}
    \centering
    \includegraphics[width=1\linewidth]{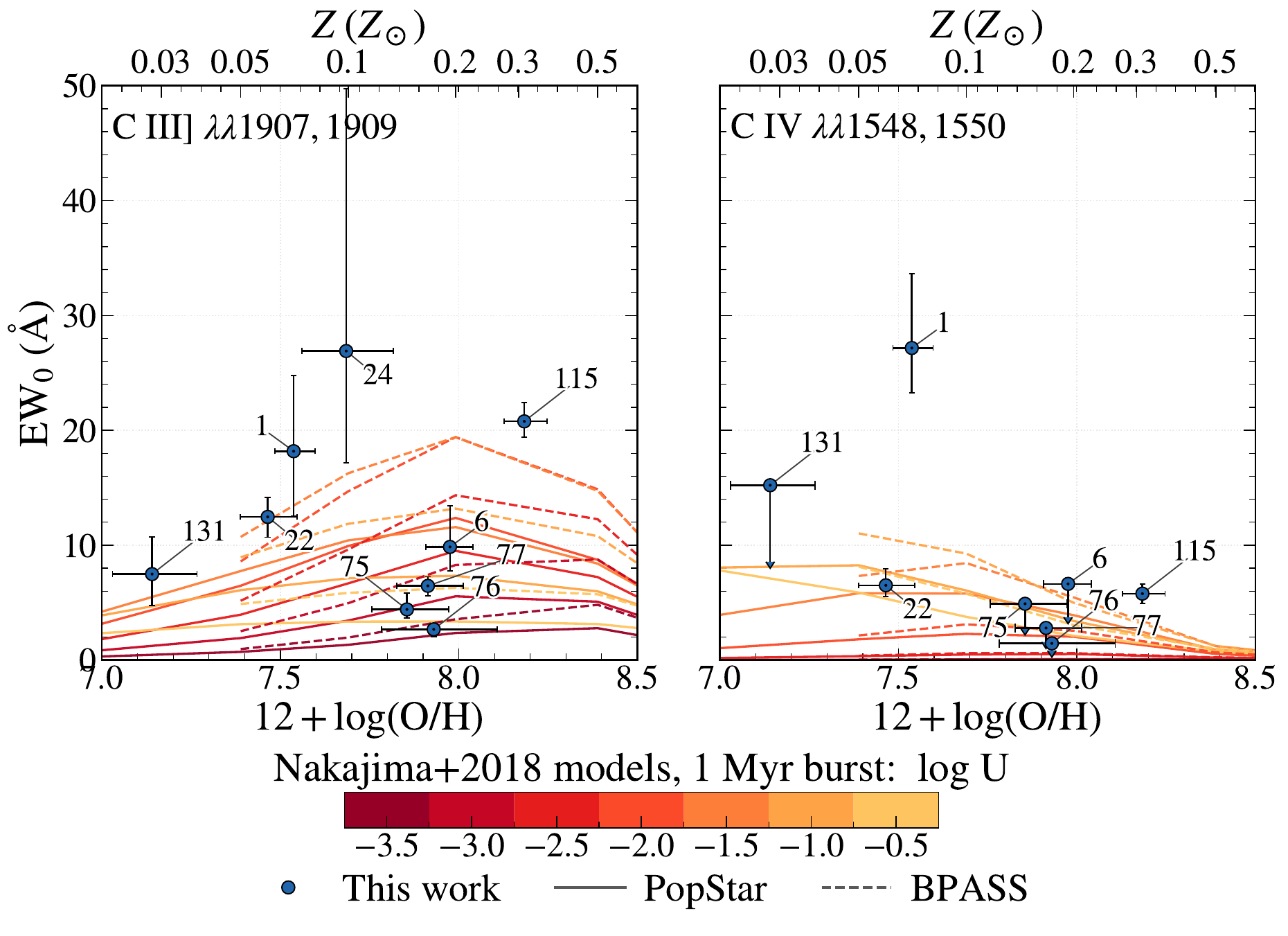}
    \caption{Rest-frame equivalent widths of C\,{\sc iii}]\,$\lambda\lambda1907,1909$ (left) and C\,{\sc iv}\,$\lambda\lambda1548,1550$ (right)  against the direct $T_{\rm e}$-based ([\oiii]$\lambda$4363) metallicity. 
    Curves are the star-forming photoionisation grids of \citet{nakajima2018} at a burst age of $1$\,Myr and $\log(n_{\rm H}/\mathrm{cm^{-3}}) = 3$. Solid curves represent the POPSTAR single-star models \citep{molla2009}, and dashed curves indicate the BPASS binary models \citep{eldridge2017}, coloured by ionisation parameter as shown in the colour-bar.}
    \label{fig:carbon_ew_nakajima}
\end{figure}

Figure~\ref{fig:carbon_ew_nakajima} shows the rest-frame equivalent widths of C\,{\sc iii}]\,$\lambda\lambda1907,1909$ \AA\ and C\,{\sc iv}\,$\lambda\lambda1548,1551$ \AA\ as a function of metallicity, compared to the photoionisation grids of \citet{nakajima2018}.
$\mathrm{EW}_0(\mathrm{C\,\textsc{iii}]})$ spans $2.7$ \AA\ to $26.9$\,\AA{} with no obvious correlation with oxygen abundance. Five of the nine sources (OMEGA-1, 22, 24, 115 and 131) nonetheless exceed the strongest C\,{\sc iii}] strengths reached by the photoionisation grids at fixed metallicity, assuming a $1$\,Myr burst at $\log(n_{\rm H}/\mathrm{cm^{-3}}) = 3$ and considering the most extreme ionisation parameters and stellar populations. Such discrepancies may point to both the more extreme nature of high-redshift galaxies compared to previous assumptions, as well as incomplete physical frameworks incorporated by such theoretical models which may require the incorporation of e.g., harder ionising spectra at fixed nebular metallicity to match such strong carbon line emitters.

At the same time, we note that the grids have a characteristic upper envelope shape, which increases in EW$_0$(C\,{\sc iii}]) as metallicity decreases, in the regime where 12+$\log({\rm O/H})\geq8.0$. At lower metallicities, the upper envelope {\it decreases} with decreasing metallicity. This shape reflects the changing C$^{2+}$/C fraction due to primarily variations of log\,$U$ at the high metallicity end, and a shift towards more pristine sources at the lower end. We note that the 3 lowest-metallicity galaxies (OMEGA-131, 22, and 1) all have C\,{\sc iii}] equivalent widths that are consistent with or lower than that of OMEGA-24, which has the highest  C\,{\sc iii}] equivalent width in the sample, at close to the median metallicity. A larger sample is required to conclusively determine if the objects presented here are detecting the theoretical decrease in C\,{\sc iii}] as metallicity decreases to the lowest values.

C\,{\sc iv} is detected in only three galaxies (OMEGA-1, 22 and 115) while remaining an upper limit among the rest. OMEGA-1 is extreme by this measure: at $\mathrm{EW}_0 = 27^{+6}_{-4}$\,\AA{} (with $S/N=15$), it exceeds the maximum equivalent width predicted by the \citet{nakajima2023} models at fixed oxygen abundance by a factor of 2.7$\times$, and four times the next strongest detection in the sample. The object also represents the highest-redshift galaxy in our sample at $z = 8.28$, the second hottest at $2.30\pm0.15\times10^{4}$\,K, and one of the three galaxies below the MZR, and as such a harder ionising spectrum relative to its \civ-weaker counterparts is consistent with the temperature and metallicity probed. While one may be tempted to claim the presence of an AGN in this source, the comparison of UV line ratios in Figure~\ref{fig:uv_diagnostics} and absence of any clear rest-optical AGN features appears to confidently reject such a scenario and favour a star-formation dominated system.

\subsection{Elemental abundance ratios}

\begin{figure}
    \centering
    \includegraphics[width=1\linewidth]{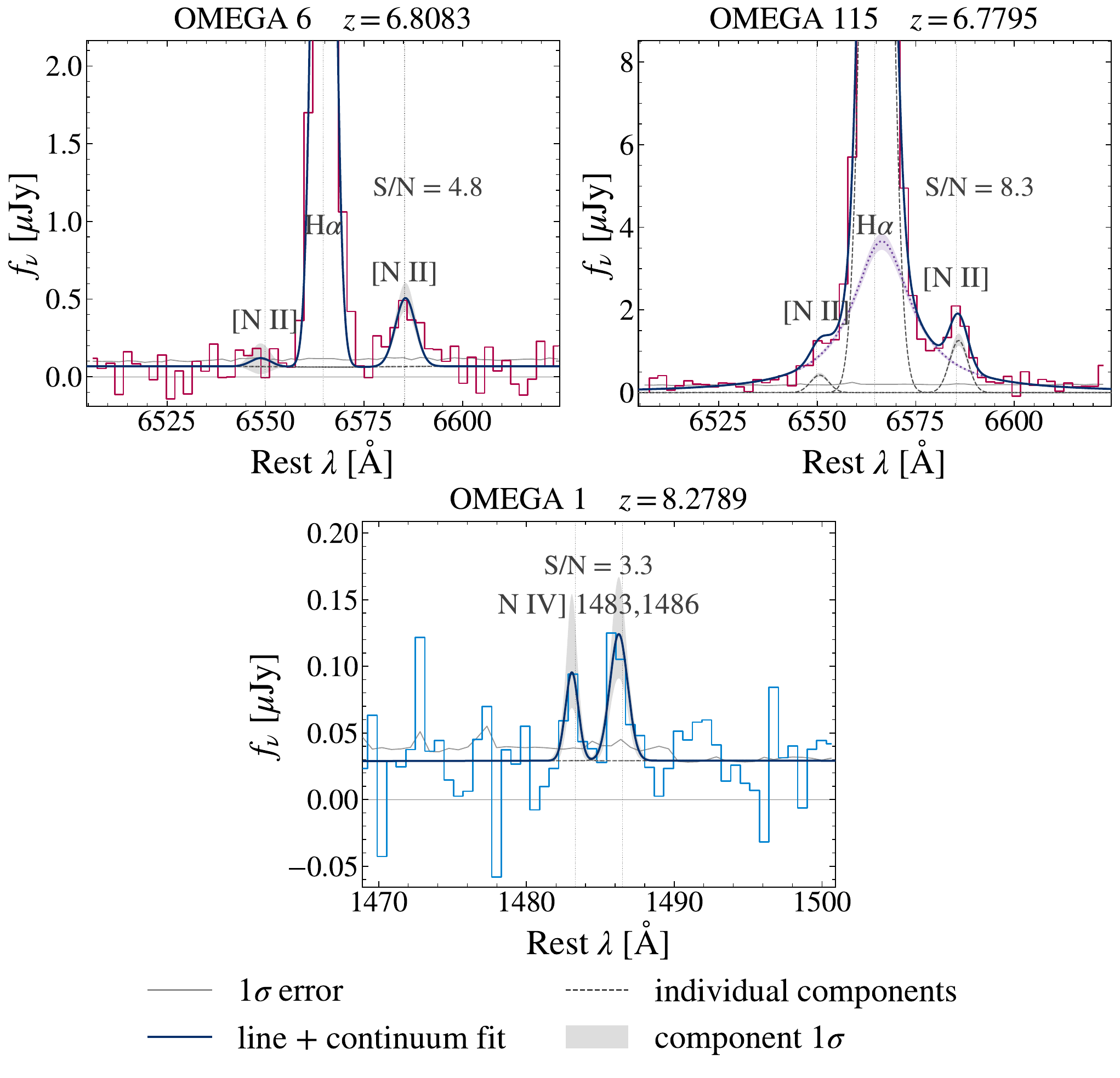}
    \caption{$>3\sigma$ detections of low- and high-ionisation nitrogen emission lines (red and blue, respectively) from which nitrogen abundances can be derived. In the case of blended profiles, both the combined (black line) and individual fits to the data are shown.}
    \label{fig:ha_nii_fits}
\end{figure}

\begin{figure}
    \centering
    \includegraphics[width=1\linewidth]{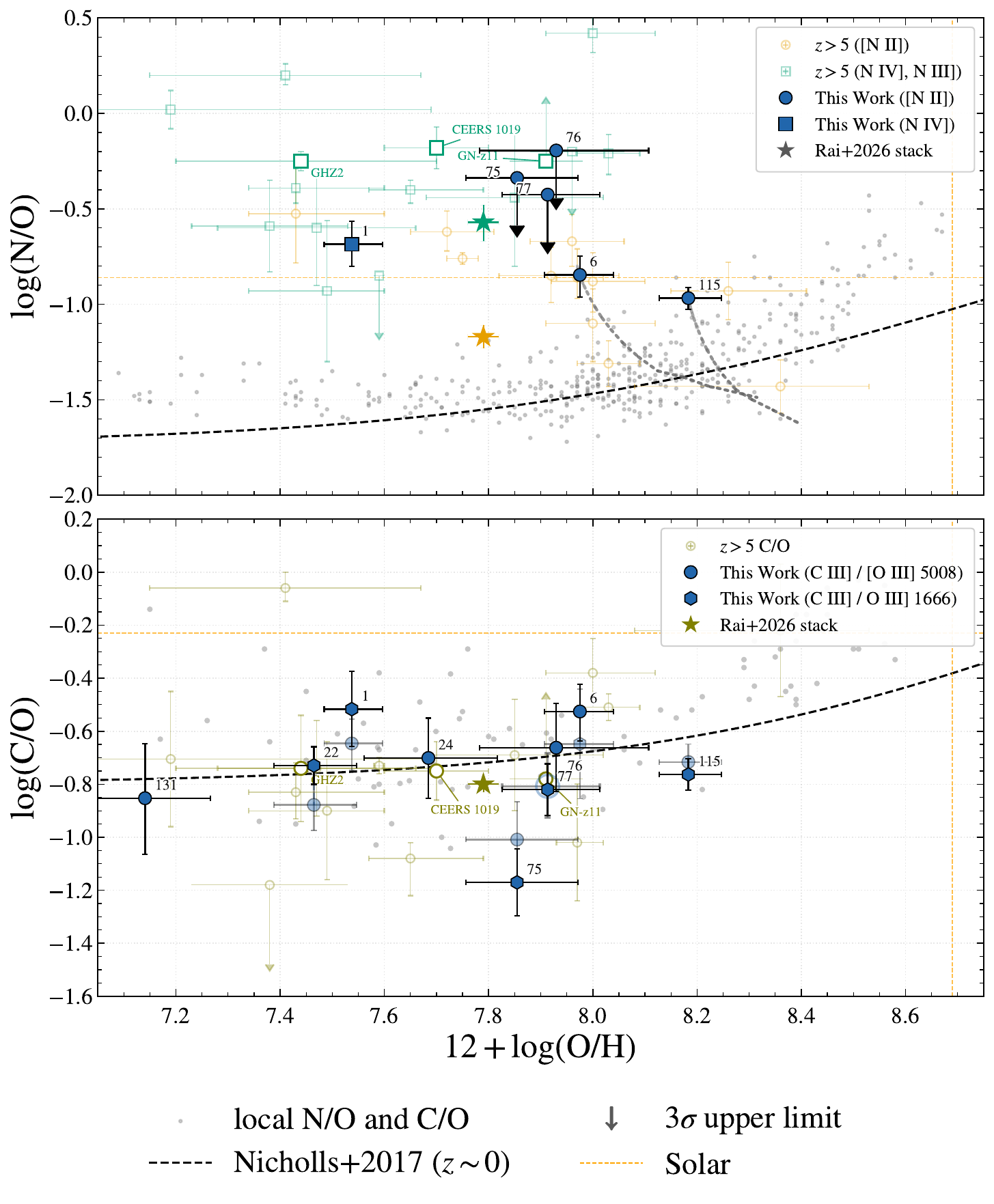}
    \caption{N/O (top) and C/O (bottom) vs. direct $T_{\rm e}$-based O/H for the nine galaxies in our sample.
    Blue points are this work, their shape naming the lines each ratio rests on: in the upper panel a circle for N/O from [\nii]\,$\lambda6585$ and a square for N/O from \niv]\,$\lambda\lambda1483,1486$, and in the lower panel a circle for C/O from \ciii]\,$\lambda\lambda1907,1909$ against [\oiii]\,$\lambda5008$ and a hexagon for C/O against \oiii]\,$\lambda1666$. Downward arrows are $3\sigma$ upper limits.
    Open symbols are individual galaxies from the literature with direct $T_{\rm e}$-based abundances, split in the upper panel by the lines their N/O rests on, [\nii] in amber and \niv], \niii] in teal.
    Grey points are local \hii\ regions above and the local C/O compilation below, the black dashed curves are the relations of \citet{nicholls2017}, and the orange dashed lines mark the solar values.
    Background coloured points  show individual high-redshift galaxies with direct $T_{\rm e}$-based abundances from the literature \citep{james2009, berg2016, vanzella2022, cameron2023, isobe2023, ubler2023, cameron2024, calabro2024, castellano2024, ji2024, labbe2024, marques2024, schaerer2024, topping2024, alvarezmarquez25, arellanocordova2025, curti2025, napolitano2025, navarrocarrera2025, stiavelli2025, tacchella2025, topping2025, welch2025, zavala25, zhang2025}. Stars mark the $z=6$-$10$ stack of \citet{rai2026}, its ultraviolet and optical N/O drawn at a common metallicity in those same two colours and joined by a thin stem, and its C/O as a single point below.
    Grey curves in the upper panel trace how each galaxy moves as the assumed low-ionisation density is varied from $10^{2}$ to $10^{5}$\,cm$^{-3}$, darkening as the density rises, and are drawn only for the two galaxies whose N/O is a detection from [\nii] rather than a limit.}
    \label{fig:no_co}
\end{figure}

With robust $T_{e}-$based oxygen abundances in hand, we now turn to comparing N and C abundances relative to O. Two OMEGA sources (OMEGA-6 and OMEGA-115) have [\nii]\,$\lambda6585$ \AA\ detected at $>3\sigma$, allowing for low-ionisation N abundance determinations, while one source (OMEGA-1) instead shows a $3.2\sigma$ \niv]\,$\lambda\lambda1483,1486$ \AA\ detection for high-ionisation determinations. We show the spectra of these detections in Figure~\ref{fig:ha_nii_fits}. For the remaining sources, we determine upper limits on N/O from the upper limits on [\nii]. Both sets of abundance ratio measurements ([\nii]- and \niv]-derived) are shown in Figure~\ref{fig:no_co}, along with a number of well-known UV-based nitrogen emitters at $z>9$ (GHZ2, GN-z11, and CEERS-1019; \citealt{castellano2024,bunker2023gnz11,marques2024}) and the composite results of \citet{rai2026}, for comparison.

The three detections span $\log(\mathrm{N/O}) = -0.97$ to $-0.69$, all of which sit close to the solar value of $-0.86$. Nitrogen-enhancements of this kind are no longer a rarity at these redshifts, although the exact prevalence remains uncertain: the number of nitrogen-rich galaxies have been accumulating steadily since the launch of JWST \citep[e.g.][]{topping2025,rusakov2026,morel2026}, while \citet{rai2026} find \niv]-based enhancement to be ubiquitous in $6 < z < 10$ galaxies through stacking of $R\sim1000$ NIRSpec spectra.

At approximately fixed metallicity, the [\nii]-derived values are consistent with others reported in the literature for individual $z\sim5$ galaxies (e.g., \citealt{arellanocordova2025}), and extend such measurements to larger samples of individual galaxies at $z>6$. Noticeably, however, and as pointed out in \citet{arellanocordova2025}, \citet{umeda2026} and \citet{rai2026}, the [\nii]-derived values lie systematically below \niv]- or \niii]-derived ones, by up to $\sim0.5-1$ dex. Although we cannot directly compare such an offset in any of our individual sources due to the lack of detections both line detections, the difference is further consolidated when comparing abundances derived from the $R\sim1000$ composite spectrum of \citet{rai2026}, which shows a similar $\sim0.6$ dex difference between the choice of tracer ($\log(\mathrm{N/O})=$ using \niv] and $\log(\mathrm{N/O})=$ using [\nii]). While a number of considerations -- such as the choice of ICF and density assumptions -- make one-to-one comparisons challenging, the offset is noteworthy and points to a likely mixture of differing ISM conditions and biases resulting from the choice of tracer and assumptions. To this end, the assumed electron density represents a fundamental consideration given the more than order-of-magnitude lower critical density of the [\oii] line ($n_{\rm crit}\sim10^{3.5}\,{\rm cm^{-3}}$) compared to that of the  [\nii] line ($n_{\rm crit}\sim10^{5}\,{\rm cm^{-3}}$), which can lead to preferential collisional de-excitation of the [\oii] line. To illustrate this effect, for the two [\nii]-detected sources, we show the impact of assumed electron density on the inferred N/O ratio. As the assumed electron density increases from $\sim$10$^{2}$ cm$^{-3}$ to 10$^{5}$ cm$^{-3}$, the inferred N/O systematically decreases (i.e., because a greater correction for collisional de-excitation is required for [\oii] than [\nii]). Although we do not have direct determinations of electron densities in low-, medium-, and high-ionisation zones for most of our targets, the detection of the [\sii]\,$\lambda\lambda6718,6732$ line pair in OMEGA-115 suggests a low-ionisation electron density of $n_{\rm e} = 95$\,cm$^{-3}$, well below the extreme values where collisional de-excitation becomes important.

OMEGA-1 is the only source with a \niv]-based N/O ratio, which resides at supersolar levels, albeit at $\sim0.5$ dex lower metallicity than its [\nii]-detected counterparts. While somewhat decreased compared to the N/O levels seen at higher metallicity, OMEGA-1 nonetheless sits at comparable values seen in other high-redshift objects at similar metallicities. Notably, its N/O lies considerably and systematically lower ($\sim0.45$ dex) than the extreme values seen in the ultra-luminous $z>9$ \niv] emitters (GHZ2, GN-z11 and CEERS-1019), although its \niv] emission is comparable in EW to that of GN-z11 ($\mathrm{EW}_0 = 4.9^{+2.4}_{-1.8}$\,\AA\ versus $5.1\pm0.9$\,\AA) and its \civ\ is stronger by almost an order of magnitude ($27.2\pm5.0$\,\AA\ versus $3.5\pm1.2$\,\AA), as reported by \citet{bunker2023gnz11}. The same comparison applies to OMEGA-6 and 115, the two sources with optical N/O measurements. At $\log(\mathrm{N/O})=-0.85$ and $-0.97$, both values are approximately solar and $+0.4-0.6$ dex above the local \citet{nicholls2017} relation at their metallicities, however they lie $0.6-0.8$\,dex below the extreme UV-based emitters, despite being moderately more metal-rich. 

In a similar fashion,  C/O is measured in all nine galaxies in our sample. The derived values range from $\log(\mathrm{C/O}) = -1.17$ to $-0.52$ with a median of $-0.73$, within $0.03$\,dex of the local \citet{nicholls2017} relation and in broad agreement with other high redshift sources and the composite spectrum of \citet{rai2026}, who report $\log(\mathrm{C/O}) = -0.80$. Eight of the nine galaxies lie below $12+\log(\mathrm{O/H}) = 8.0$, identified by \citet{berg2019} as the plateau of the \citet{nicholls2017} relation, where carbon is produced by primary, metallicity-independent nucleosynthesis at $\log(\mathrm{C/O}) = -0.8$ and the pseudo-secondary contribution from the metallicity-driven winds of massive stars \citep{henry2000} has not yet set in.

\section{Discussion}
\label{sec:discussion}

In the previous sections we have highlighted the power of ultra-deep spectroscopy across both rest-UV and rest-optical wavelengths. Such combined datasets unveil fainter diagnostic features that greatly improve the robustness of ISM, stellar, and non-thermal characterisations, and allow for a more comprehensive multi-wavelength analysis. In this respect, one of the main results to emerge from this analysis is the detection of nitrogen lines (either [\nii] or \niv]) leading to both supersolar N/O abundance ratios and systematically lower values of N/O based on rest-optical diagnostics compared to rest-UV ones. The former is consistent with studies of both individual and composite spectra (e.g., \citealt{topping2025,arellanocordova2025,rai2026}), while the latter highlights the dangers and systematics associated with using a single nitrogen ionisation state to infer the degree of nitrogen enrichment in the first billion years.

There is an added complexity in the rest-optical-based N/O values, in that the critical densities for both [\nii] and [\oii] features are orders of magnitude lower than those of their higher-ionisation, rest-frame UV counterparts (i.e., \niii], \niv], \oiii]). In gas whose density rivals the critical density of [\oii]
($\sim10^{3.5}\,{\rm cm^{-3}}$) or even that of [\nii] ($n_{\rm crit}\sim10^{5}$), the effects of collisional de-excitation must be accounted for when estimating N/O from rest-optical tracers. Another relevant factor is that [\nii] may become undetectably faint relative to higher-ionisation species \citep{shapley2023} in high-ionisation gas typical of low-metallicity galaxies at $z>6$. However, if the correct physical conditions are assumed (e.g., density, ionisation parameter, ionising spectrum), the same N/O should be recovered regardless of which tracer (rest-UV or rest-optical) is used. A larger sample of $z>6$ galaxies with coverage of both rest-UV and rest-optical nitrogen and oxygen lines, probing multiple ionisation stages, is clearly needed to understand this problem. A larger sample of direct measures of electron density is also essential for addressing this problem.

Despite these important systematics based on different nitrogen tracers, we note that the majority of high-EW \niv] ($\gtrsim$5-10 \AA; \citealt{morel26}) has been found in ultra-luminous ($M_{\rm UV}\lesssim-21$) sources \citep{bunker2023gnz11,marques2024,castellano2024}, while moderately luminous ($-21\lesssim M_{\rm UV}\lesssim-19$) sources display far weaker EWs ($\sim1$ \AA; \citealt{rai2026}), possibly indicating a connection between different star-formation conditions in the most luminous samples compared to the wider population of star-forming sources at high redshift. Indeed, a generic but intensely bursty star-forming origin was recently proposed by \citet{robertsborsani2026} to explain an association between strong \niv] and \civ\ emission in the most luminous $z>10$ galaxies. Interestingly, we note that the only source in our sample with a \niv] detection (OMEGA-1) happens to display the strongest \civ\ EW, supporting this argument.

Our sample generally has high $O32$ ratios $\gtrsim10$, implying hard radiation fields, high ionisation parameters, and/or high gas densities, all of which suppress emission from low-ionisation [\oii].
Among the sources in Figure~\ref{fig:no_co}, the two detected [\nii]-emitters OMEGA-6 and OMEGA-115 have lower ratios ($O32=8.5^{+0.8}_{-1.2}$ and $11.3^{+3.2}_{-2.0}$) than the nitrogen-undetected sources OMEGA-75, 76, and 77 ($O32\simeq12.9$--$14.2$).
Given the higher critical density of [\nii] than [\oii], this difference suggests that the higher $O32$ values in the [\nii]-undetected sources may be driven by a higher ionisation state rather than high densities, since the latter suppresses only [\oii] at $n_e\sim10^3-10^5$~cm$^{-3}$.
OMEGA-1, the only source with a \niv] detection and no [\nii] counterpart, lies well above the rest with a 3$\sigma$ lower limit of $O32>77$, and also displays high-EW \civ\ and a large \civ/\ciii] ratio, all suggestive of a very high degree of ionisation.
Its extreme $O32$ ratio is more easily explained by the combination of very high gas density and ionisation state, and would require an extremely hard incident spectrum and/or extremely large ionisation parameter at low density.
Significant improvements on inferences of the ionisation parameter and ionising spectral shape via photoionisation modelling can be made when the gas-phase metallicity and abundance pattern are known, eliminating degrees of freedom in the models (e.g., varying C/O in the models at fixed O/H and ionisation parameter directly modulates the \ciii] and \civ\ EWs; Figure~\ref{fig:carbon_ew_nakajima}).
The powerful combination of deep rest-UV and rest-optical spectroscopy demonstrated in this analysis provides precise constraints on O/H, C/O, and N/O that can in turn lead to an improved understanding of the ionisation state and ionising sources in these systems.

The absence of strong evidence for AGN and first-order trends with density and/or radiation field arguments suggest star formation origins driving enhanced N/O, as has been suggested by a number of recent works at high redshift (e.g., \citealt{marques2024,marqueschaves2026,robertsborsani2026,rai2026,berg2026}).
The scatter in C/O ratios around the local relation measured across our sample (Figure~\ref{fig:no_co}) offers a further handle on the origin of the nitrogen enrichment.
IMF-integrated core-collapse supernova yields predict $\log(\mathrm{C/O})\approx-0.8$ to $-1.1$ \citep{kobayashi2006,tominaga2007,nomoto2013,jones2023}.
Seven of our nine sources have C/O consistent with those yields, as expected if their carbon and oxygen come from core-collapse supernovae alone, while OMEGA-1 and -6 sit within $2.5\sigma$ of the same range. An elevated N/O accompanied by normal C/O with low scatter can be interpreted as the signature of enrichment by CNO-processed, hot hydrogen-burning material, in which carbon and oxygen are converted into nitrogen while C/O remains largely unchanged. This pattern favours nucleosynthesis in supermassive stars or AGB stars, as invoked for strong \niv]-emitters such as GN-z11, CEERS-1019, and GHZ2 \citep{bunker2023gnz11,marques2024,castellano2024}.
Our rest-optical spectra show no detected signatures of Wolf-Rayet stars, including the possible presence of emission line ``bump'' complexes (neither the blue bump from \heii$\,\lambda4686$ with C\,{\sc iii}]/\civ$\,\lambda\lambda4650$, nor the one with \civ$\,\lambda5808$).
However, even with our deep G395M integrations, we cannot rule out the presence of Wolf-Rayet stars that may contribute to N-enhancement due to the extreme faintness of these Wolf-Rayet features in an integrated stellar population, which have been detected in a N-enhanced source at $z\sim6$ leveraging the combination of extremely deep NIRSpec integrations and gravitational lensing \citep{berg2026}.

\section{Summary and Conclusions}
\label{sec:summary}

We have presented ultra-deep JWST/NIRSpec $R\sim1000$ spectroscopy of nine star-forming galaxies at $z=6.0-8.3$ in GOODS-N, combining rest-optical G395M observations from the OMEGA survey (GO 7729; PI Roberts-Borsani) with ancillary rest-UV G140M spectroscopy from the SPURS, DIVER, AURORA, JADES and GO-4762 programmes. The resulting continuous rest-frame UV-to-optical coverage provides, for each galaxy, individually, the electron temperature, density and ionisation correction factors required for precise, direct-method C-N-O abundances.
These are moderately bright ($M_{\rm UV}=-18.67$ to $-21.1$), unlensed galaxies with no AGN signature in any rest-UV diagnostic, sitting towards the upper end of the star-forming main sequence as their emission-line selection implies. They are, in other words, broadly representative of the actively star-forming population during the reionisation era, in a regime where abundance measurements of individual objects have until now demanded strong lensing, exceptional luminosity, or stacking.
Our key results are as follows.

\begin{enumerate}

    \item The auroral {[\oiii]}\,$\lambda$4363 \AA\ line is detected in all nine galaxies at S/N $=3.3-10.8$, yielding $T_{\rm e}=1.27-2.42\times10^{4}$\,K and direct-method abundances of $12+\log(\mathrm{O/H})=7.14-8.18$ without recourse to strong-line calibrations or stacking.

    \vspace{0.2cm}
    
    \item Beyond the \ciii]$\lambda\lambda$1907,1909 \AA\ and [\oiii]\,$\lambda$4363 \AA\ detections on upon which the present sample is selected, the significant depth of the spectra reveals features rarely accessible in individual galaxies at these redshifts: \civ\ emission in three galaxies, \heii\,$\lambda$1640 in four, \heii\,$\lambda$4686 in OMEGA-115, and both \civ\ and \nv\ P-Cygni profiles in OMEGA-76. Interstellar absorption is detected in both ionisation phases, the low ionisation phase in three galaxies and the high ionisation phase in two. We observe kinematically distinct phases: in OMEGA-76 and -77 the low ionisation phase is consistent with systemic velocities within $1\sigma$, while the high ionisation gas is blueshifted by $181\pm51$ and $192\pm29$\,\kms respectively, suggesting faster and more ionised outflows.
    
    \vspace{0.2cm}

    \item C/O is measured in all nine galaxies in our sample, spanning $\log(\mathrm{C/O})=-1.17$ to $-0.52$ (median $-0.73$) and sitting on the local \citet{nicholls2017} relation at the plateau where carbon production is primary. N/O is measured in three galaxies, from $\log(\mathrm{N/O})=-0.97$ to $-0.69$ using both ultraviolet (\niv]) and optical ([\nii]) tracers, elevated relative to local samples at approximately solar values, but below the especially high values seen in the extreme nitrogen emitters GN-z11, CEERS-1019, and GHZ2. In the three galaxies where [\nii]\,$\lambda$6585 \AA\ is covered but falls below our S/N threshold, we place stringent $3\sigma$ upper limits on N/O, each consistent with the enhancement measured in the three N detections.

    \vspace{0.2cm}

    \item N/O derived from optical tracers is density dependent at fixed [\nii]/[\oii] line ratio. If the actual density our sample is higher than the one assumed, our inferred N/O values are biased high relative to the true ones. We show that the apparent N/O enhancement disappears entirely if the density is significantly higher than assumed. Specifically, OMEGA-6 and 115 reach the local N/O vs. O/H relation at densities of $3.3\times10^{4}$ and $1.1\times10^{4}$\,cm$^{-3}$ respectively. Detection of the [\sii]\,$\lambda\lambda6718,6732$ \AA\ doublet in OMEGA-115, however, suggests electron densities of $\sim95$ $cm^{-3}$ in the low-ionisation zone, far below such extreme values and suggestive of true nitrogen enhancement in these sources.

\end{enumerate}

The unprecedented power of JWST spectroscopy has now enabled characterisation of galaxy properties and their underlying chemical enrichment out to the earliest times. Ultra-deep and $R\gtrsim1000$ NIRSpec spectroscopy over rest-frame UV-to-optical wavelengths remains a prerequisite to this endeavour, given its ability to recover faint and blended ISM, stellar, and SMBH signatures that lie beyond the constraining power of the vast majority of archival spectra to date. In this pilot study, we have shown the remarkable diagnostic tools and power enabled by such spectroscopy over individual $z>6$ sources, allowing for accurate characterisations of their underlying properties, chemical enrichment, and potential biases. A key next step will be to extend such observations to larger samples, enabling population-level studies with which to obtain a census on the radiation fields, enrichment histories, and SMBH contributions that dominated the earliest stages of galaxy formation and evolution after the Big Bang.

\section*{Acknowledgements}

RSR acknowledges support from the Science and Technology Facilities Council (STFC) through a PhD studentship under grant number UKRI1789. Data Reduction was performed using the UCL Myriad High Performance Computing Facility (Myriad@UCL) and associated support services.

This work is based on observations made with the NASA/ESA/CSA James Webb Space Telescope, obtained from the Mikulski Archive for Space Telescopes at the Space Telescope Science Institute, which is operated by the Association of Universities for Research in Astronomy, Inc., under NASA contract NAS 5-03127 for JWST.
Support for Program number JWST-GO-07729 was provided through a grant from the STScI under NASA contract NAS 5-03127.

This work has received funding from the Swiss State Secretariat for Education, Research and Innovation (SERI) under contract number MB22.00072, as well as from the Swiss National Science Foundation (SNSF) through project grants 200020\_207349 and 2000-1-243073. The Cosmic Dawn Center (DAWN) is funded by the Danish National Research Foundation under grant DNRF140.

\section*{Data Availability}

The JWST/NIRSpec data from programmes 9214, 8018, 1914, 1181 and 4762 are publicly available from the Mikulski Archive for Space Telescopes (MAST; \url{http://archive.stsci.edu}).
Data from the OMEGA survey (GO 7729) are under exclusive access and will be accessible from MAST upon expiration of the exclusive access period. All data products presented in this paper are available upon reasonable request to the authors.



\bibliographystyle{mnras}
\bibliography{references} 




\appendix
\section{Source morphologies and derived ISM conditions}

\begin{figure}
    \centering
    \includegraphics[width=1\linewidth]{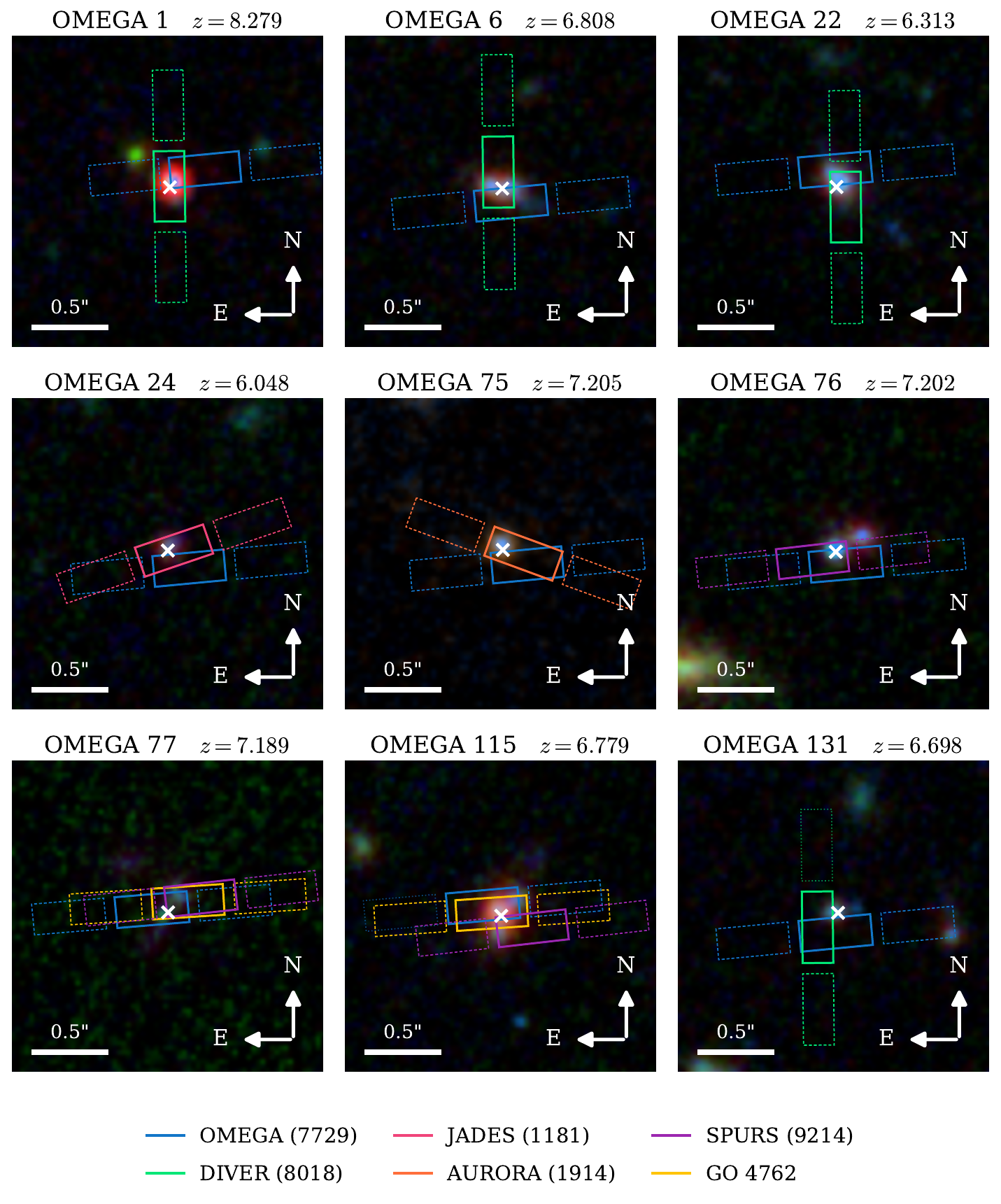}
    \caption{JWST/NIRCam colour composites of the nine star-forming OMEGA galaxies, cut from the JADES GOODS-N mosaics \citep{eisenstein2023,rieke2023}. Red, green and blue are F444W, F200W and F115W respectively, with F150W substituted for F200W in OMEGA~77, which F200W is not available. OMEGA~75 is covered by neither F200W nor F150W, and its green channel is interpolated from the other two. Overlaid are the NIRSpec micro-shutter footprints of every programme that observed the source.}
    \label{fig:cutouts}
\end{figure}

\begin{table*}
\centering
\small
\setlength{\tabcolsep}{4pt}
\renewcommand{\arraystretch}{1.35}
\begin{tabular}{lccccccccccc}
\toprule
OMEGA & $A_V$ & $12+\log(\mathrm{O/H})$ & $\log(\mathrm{C/O})$ & $\log(\mathrm{N/O})$ & $n_{\rm e}^{\rm high}$ & $n_{\rm e}^{\rm mid}$ & $n_{\rm e}^{\rm low}$ & $T_{\rm e}^{\rm high}$ & $T_{\rm e}^{\rm mid}$ & $T_{\rm e}^{\rm low}$ & $\log U$ \\
MSAID & (mag) & & & & ($10^4$\,cm$^{-3}$) & ($10^4$\,cm$^{-3}$) & (cm$^{-3}$) & ($10^4$\,K) & ($10^4$\,K) & ($10^4$\,K) & \\
\midrule
1 & $0.39 \pm 0.37$ & $7.54^{+0.06}_{-0.05}$ & $-0.52^{+0.14}_{-0.14}$ & $-0.69^{+0.12}_{-0.11}$ & $22.2^{+26.4}_{-16.9}$ & $29.13^{+61.56}_{-18.17}$ & $782^{*}$ & $2.30 \pm 0.15$ & $2.08 \pm 0.13$ & $1.91 \pm 0.11$ & $-1.04^{+0.03}_{-0.06}$ \\
6 & $0.17 \pm 0.07$ & $7.98^{+0.06}_{-0.07}$ & $-0.53^{+0.10}_{-0.11}\,^{\ddagger}$ & $-0.85^{+0.10}_{-0.11}$ & $15.1^{*}$ & $4.91^{+5.11}_{-2.87}$ & $636^{*}$ & $1.46 \pm 0.09$ & $1.38 \pm 0.07$ & $1.32 \pm 0.06$ & $-1.68^{+0.25}_{-0.19}$ \\
22 & $0.00 \pm 0.05$ & $7.46^{+0.08}_{-0.08}$ & $-0.73^{+0.07}_{-0.07}$ & -- & $13.6^{*}$ & $0.13^{+0.82}_{-0.13}\,^{\dagger}$ & $588^{*}$ & $1.97 \pm 0.17$ & $1.81 \pm 0.14$ & $1.68 \pm 0.12$ & $-1.73^{+0.43}_{-0.34}$ \\
24 & $0.00 \pm 0.94$ & $7.68^{+0.13}_{-0.12}$ & $-0.70^{+0.15}_{-0.15}\,^{\ddagger}$ & -- & $12.8^{*}$ & $4.42^{+7.64}_{-3.09}$ & $562^{*}$ & $1.97 \pm 0.29$ & $1.81 \pm 0.24$ & $1.68 \pm 0.20$ & $-1.83^{+0.18}_{-0.23}$ \\
75 & $0.00 \pm 0.13$ & $7.85^{+0.12}_{-0.10}$ & $-1.17^{+0.13}_{-0.13}$ & $<-0.34$ & $16.3^{*}$ & $2.29^{+3.91}_{-2.03}$ & $675^{*}$ & $1.56 \pm 0.15$ & $1.46 \pm 0.13$ & $1.39 \pm 0.11$ & $-1.77^{+0.60}_{-0.32}$ \\
76 & $0.09 \pm 0.80$ & $7.93^{+0.18}_{-0.15}$ & $-0.66^{+0.17}_{-0.16}\,^{\ddagger}$ & $<-0.19$ & $16.3^{*}$ & $1.34^{+2.66}_{-1.34}\,^{\dagger}$ & $675^{*}$ & $1.52 \pm 0.23$ & $1.43 \pm 0.19$ & $1.36 \pm 0.16$ & $-1.29^{+0.19}_{-0.49}$ \\
77 & $0.00 \pm 0.11$ & $7.91^{+0.10}_{-0.09}$ & $-0.82^{+0.10}_{-0.10}$ & $<-0.43$ & $16.3^{*}$ & $0.65^{+1.34}_{-0.65}\,^{\dagger}$ & $673^{*}$ & $1.57 \pm 0.14$ & $1.48 \pm 0.11$ & $1.40 \pm 0.10$ & $-2.46^{+0.51}_{-0.18}$ \\
115 & $0.02 \pm 0.21$ & $8.18^{+0.06}_{-0.06}$ & $-0.76^{+0.06}_{-0.06}$ & $-0.97^{+0.05}_{-0.06}$ & $15.0^{*}$ & $3.68^{+0.98}_{-0.67}$ & $95^{+754}_{-94}\,^{\dagger}$ & $1.27 \pm 0.06$ & $1.22 \pm 0.05$ & $1.19 \pm 0.04$ & $-2.52^{+0.06}_{-0.07}$ \\
131 & $0.19 \pm 0.39$ & $7.14^{+0.13}_{-0.11}$ & $-0.85^{+0.20}_{-0.21}\,^{\ddagger}$ & -- & $14.7^{*}$ & $1.71^{+6.47}_{-1.71}\,^{\dagger}$ & $625^{*}$ & $2.42 \pm 0.39$ & $2.18 \pm 0.32$ & $1.99 \pm 0.27$ & $-1.19^{+0.12}_{-0.78}$ \\
\bottomrule
\end{tabular}
\caption{Ionised gas conditions and elemental abundances for the nine OMEGA galaxies, from the framework of Section~\ref{sec:chemical_abund}.
$A_V$ is from the Balmer decrements on the \citet{cardelli1989} law at $R_V=3.1$; $12+\log(\mathrm{O/H})$ is the direct-method value anchored on [\oiii]\,$\lambda4363$.
$\log(\mathrm{C/O})$ is the adopted route, C\,{\sc iii}]\,$\lambda\lambda1907,1909$ against O\,{\sc iii}]\,$\lambda1666$ except where marked $^{\ddagger}$, which use [\oiii]\,$\lambda5008$ because O\,{\sc iii}]\,$\lambda1666$ is undetected.
$\log(\mathrm{N/O})$ is from [\nii]\,$\lambda6585$ except OMEGA-1, which uses \niv]\,$\lambda\lambda1483,1486$; entries marked $<$ are $3\sigma$ upper limits, matching Figure~\ref{fig:no_co}, and -- indicates no coverage of the required lines.
Densities come from the \niv]\,$\lambda\lambda1483,1486$, C\,{\sc iii}]\,$\lambda\lambda1907,1909$ and [\sii]\,$\lambda\lambda6718,6732$ doublets for the high-, intermediate- and low-ionisation zones respectively.
A value marked $^{*}$ is not measured and comes instead from the redshift-scaled relations $n_{\rm e}^{\rm low}=54(1+z)^{1.2}$ \citep{abdurrouf2024} and $n_{\rm e}^{\rm high}=5400(1+z)^{1.62}$ \citep{martinez2025}; $^{\dagger}$ marks an entry whose lower bound is consistent with the low-density limit of its doublet.
$T_{\rm e}^{\rm high}$ is measured from [\oiii]\,$\lambda4363$; $T_{\rm e}^{\rm mid}$ and $T_{\rm e}^{\rm low}$ follow from it through the \citet{garnett1992} relations adopted by \citet{martinez2025}, $T^{\rm mid}=0.83\,T^{\rm high}+1700$\,K and $T^{\rm low}=0.70\,T^{\rm high}+3000$\,K, with the uncertainty scaled accordingly.
$\log U$ is from the \textsc{Prospector} fits of Section~\ref{sec:sed_fitting}.}
\label{tab:conditions}
\end{table*}



\bsp	
\label{lastpage}
\end{document}